\documentclass[10pt,journal]{IEEEtran}
\usepackage{amsmath,amssymb,amsfonts,mathtools}
\usepackage{amsthm}
\usepackage{graphicx}
\usepackage{booktabs}
\usepackage{placeins}
\usepackage{epstopdf}
\usepackage{balance}
\usepackage{cite}
\usepackage{url}
\usepackage{xcolor}
\usepackage[caption=false,font=footnotesize]{subfig}
\usepackage[ruled,vlined,linesnumbered]{algorithm2e}

\graphicspath{{native_figures/}{figure_formatting/}{figures/}}

\DeclareMathOperator*{\argmax}{argmax}

\newtheorem{theorem}{Theorem}

\newcommand{\C}{\mathbb{C}}
\newcommand{\R}{\mathbb{R}}
\newcommand{\E}{\mathbb{E}}

\newcommand{\diag}{\operatorname{diag}}
\newcommand{\rank}{\operatorname{rank}}

\renewcommand{\Re}{\operatorname{Re}}

\renewcommand{\Im}{\operatorname{Im}}

\theoremstyle{remark}
\newtheorem{remark}{Remark}
\theoremstyle{plain}

\makeatletter
\let\originalIEEEcaption\@makecaption
\long\def\@makecaption#1#2{%
  \ifx\@captype\@IEEEtablestring
    \originalIEEEcaption{#1}{#2}%
  \else
    \@IEEEfigurecaptionsepspace
    {\normalfont\footnotesize
     \noindent\parbox{\linewidth}{\centering #1.\nobreakspace\nobreakspace #2\par}\par}%
  \fi}
\makeatother

\begin{document}
\title{Ray-Tracing-Based Dynamic Channel Sounding With Autoregressive Wi-Fi CSI Prediction}
\author{Lyutianyang Zhang, \IEEEmembership{Member,~IEEE}, Yixin Ying, Liu Cao, \IEEEmembership{Member,~IEEE}, Vanlin Sathya, \IEEEmembership{Member,~IEEE},\\ Jinke Ren, \IEEEmembership{Member,~IEEE}, and Shuguang Cui, \IEEEmembership{Fellow, IEEE}%
\thanks{Lyutianyang Zhang is with the School of Microelectronics and Communication Engineering, Chongqing University, Chongqing, China (e-mail: zhanglyutianyang@cqu.edu.cn). Yixin Ying and Liu Cao are with City University of Hong Kong (Dongguan), Dongguan, Guangdong, China (e-mail: \{72516005, liu.cao\}@cityu-dg.edu.cn). Vanlin Sathya is with Celona Inc. at Campbell, California, USA (e-mail:
vanlin@celona.io). Jinke Ren and Shuguang Cui are with the School of Science and Engineering, the Shenzhen Future Network of Intelligence Institute, and the Guangdong Provincial Key Laboratory of Future Networks of Intelligence, The Chinese University of Hong Kong (Shenzhen), Shenzhen, Guangdong 518172, China (e-mail: \{jinkeren, shuguangcui\}@cuhk.edu.cn). (Corresponding author: Liu Cao)}}

\maketitle

\begin{abstract}
Channel sounding in Wi-Fi multiple-input multiple-output (MIMO)
systems must balance channel state information (CSI) accuracy
against CSI acquisition overhead. Device mobility causes channel aging, while autoregressive CSI prediction introduces errors that accumulate over sounding intervals. However, standardized statistical channel models cannot fully capture the environment-specific channel dynamics encountered in physical artificial intelligence (AI) applications, motivating the use of Sionna-based ray tracing to evaluate CSI prediction under device mobility and evolving propagation geometry. This paper develops a dynamic sounding
framework that integrates autoregressive CSI prediction with
signal-to-interference-plus-noise ratio (SINR)-constrained interval
selection. A Transformer with rotary position embedding that converts device positions into CSI sequences predicts CSI across multiple subcarriers for beamforming updates between sounding exchanges.
We formulate sounding-interval selection as a predicted average-throughput maximization problem under per-stream SINR lower-bound constraints, including CSI acquisition and processing overhead.
A prediction-error perturbation analysis provides the SINR bounds,
and cumulative-sum enumeration globally solves the finite
single-variable problem. Actual SINR feasibility holds whenever
the prediction errors satisfy the bounds. Ray-tracing
experiments assess channel prediction,
trajectory transfer, and autoregressive prediction errors. In single-STA Wi-Fi 7
link-level simulations, predicted-CSI beamforming improves mean
goodput by 67.3\%, 112.7\%, and 119.1\% over stale CSI for
linear, circular, and cosine trajectories, respectively.
\end{abstract}

\section{Introduction}
\label{sec:introduction}

\IEEEPARstart{W}{i-Fi} 8, developed through IEEE 802.11bn
\cite{11bn}, targets ultra-high reliability (UHR) under challenging
network conditions~\cite{IEEE80211bnReport}. Meanwhile, research
toward a prospective Wi-Fi 9 generation explores a greater role
for network intelligence. The formation of the IEEE 802.11 WLAN
Intelligent Networking (WIN) study group and ongoing investigations
of artificial intelligence (AI) and machine learning (AI/ML) within
IEEE 802.11 reflect this direction~\cite{IEEE80211WG,IEEE80211AIML}.
These developments motivate AI/ML models that anticipate
changing radio conditions and support adaptive transmission
decisions. However, prediction accuracy alone does not determine communication performance; it is necessary to investigate how to reduce the channel sounding overhead.

In Wi-Fi multiple-input multiple-output (MIMO) systems, an access point (AP) spatially multiplexes data streams over common time--frequency resources through transmit beamforming. In downlink multi-user MIMO (DL MU-MIMO), these streams can simultaneously serve multiple stations (STAs), i.e., mobile devices, provided that the
precoders appropriately separate the intended transmissions. Accurate channel state information (CSI) is therefore essential for beamforming and interference suppression. However, STA mobility changes the propagation channels after
sounding, creating a mismatch with the previously computed beamforming matrices. This mismatch can weaken the desired signals and introduce residual interference. More frequent sounding refreshes
CSI but consumes training and feedback
airtime~\cite{Hassibi2003Training,Caire2010TrainingFeedback}. Moreover, the feedback burden depends on the number of spatial streams, subcarriers grouping, and etc., while the associated frame exchanges and interframe spaces for CSI acquisition further reduce
airtime efficiency~\cite{Zhang2026Sounding}. Thus, channel sounding-interval optimization must balance physical (PHY)-layer transmission quality against the medium access control (MAC)-layer airtime available for the transmitted data.

Channel prediction provides a means to alleviate this tradeoff by estimating future CSI from historical observations, allowing beamforming matrices to be updated between sounding exchanges. Classical
prediction and channel-aging analyses characterize temporal channel variation and the consequences of delayed CSI~\cite{Schafhuber2005Prediction,Truong2013Aging}.
Angular-delay and propagation parameter methods exploit multipath
structure~\cite{Yin2019PAD,Qin2022PartialReciprocity}, whereas
spatial-temporal predictors exploit correlations among channel
coefficients~\cite{Wu2021STARPredict} on one subcarrier in cellular networks based on perfect CSI feedback. In~\cite{9676455}, a convolutional neural network and a convolutional long short-term
memory are combined for single-step and multi-step channel
prediction in high-speed railway networks. Spatial and temporal
correlations are learned from propagation-graph-generated channel
data, illustrating the value of incorporating propagation
structure into prediction rather than treating channel
coefficients as unrelated time series.

Transformer-based channel prediction uses attention to capture
dependencies in historical CSI~\cite{Jiang2022TransformerPredict},
while alternative architectures, such as LinFormer, further
investigate prediction accuracy and computational
efficiency~\cite{Jin2025LinFormer}. Continuous-time, pretrained,
and autoregressive approaches extend the temporal and
representation choices available for CSI
prediction~\cite{Sang2026Continuous,Liu2024LLM4CP,Fan2024CsiLLM,Li2025CSIAlignment}. As a related application, the large language model (LLM)-based method in~\cite{10892257} predicts future beam indices from
historical beam indices and angles of departure. Such beam
selection differs from predicting the complex channel matrices
used here for singular value decomposition (SVD)-based precoding
and evaluation of inter-stream interference. The comparison
in~\cite{stenhammar2024comparison} further examines neural-network-based
predictors over different prediction horizons, considering
prediction quality and implementation complexity. These studies
motivate learned prediction, but sounding control additionally
requires relating the prediction horizon to transmission
requirements and CSI acquisition overhead.

As more related work on Wi-Fi sounding,
\cite{Nabetani2018Sounding} investigates sounding-protocol design,
while~\cite{jingyuan} implements the IEEE 802.11ax channel-sounding
frame exchange in ns-3 and evaluates its overhead. The latter
provides MAC-layer support for assessing CSI acquisition overhead,
rather than a prediction-error-constrained sounding controller.
In~\cite{Ma2017Dynamic}, a low-complexity dynamic sounding method
adaptively adjusts the sounding interval in IEEE 802.11ac
MU-MIMO networks using information obtained during operation.
The impact of the explicit sounding period on Wi-Fi MU-MIMO
performance is also investigated
in~\cite{endovitskiy2024impact}, while feedback-rate selection
addresses another component of CSI acquisition
overhead~\cite{Lee2024FeedbackRate}. These approaches establish the
importance of sounding overhead and interval selection while our
focus is determining how long the predicted CSI can support
transmission when its errors affect individual spatial streams.

Learning-based CSI acquisition offers a complementary approach.
DeepMux~\cite{9448323} combines AP-side learned channel sounding
with resource allocation for joint MU-MIMO and orthogonal
frequency-division multiple access (OFDMA) transmission in IEEE
802.11ax. It reconstructs frequency-domain CSI from feedback
on a subset of tones and uses a neural network for resource-unit
assignment, addressing both reporting overhead and resource
allocation. In cellular massive MIMO,
\cite{10476328} uses a channel predictor at the base station and
a lightweight predictor function at the user equipment to reduce
feedback overhead without requiring a learned model. These works show that learned acquisition and prediction can reduce CSI-related costs, which motivate our investigation of the deployment of light Transformer at AP side for CSI prediction based on imperfect CSI with information compression and additional noise, so as to eliminate the energy consumption at the STA side while completely aligning with the IEEE channel sounding protocol. 

Related cross-layer Wi-Fi studies investigate the joint coordinated-beamforming
and cluster formation problem in WiFi 8 multi-AP Coordination scenario~\cite{Zhang2026CoBF} and optimization toward multi-link operation 
in terms of low-latency ~\cite{Zhang2026MLOScheduling} and AP--STA association and link allocation~\cite{Zhang2024MLOThroughput}. Our earlier
work~\cite{Zhang2026Sounding} specifically develops a hybrid
model-driven and data-driven sounding framework for IEEE
802.11be downlink MIMO CSI prediction on a single subcarrier. It combines MAC-layer sounding airtime with PHY-layer channel aging to optimize the average
data rate, using IEEE 802.11be Task Group (TGbe) channels and
a first-order autoregressive, AR(1), aging model. An
encoder--decoder Transformer predicts future beamforming
information from historical partial CSI feedback, with
interpolation providing finer-grained beamforming updates.

In summary, the proposed framework in this paper
considers motion-dependent ray-traced channels and autoregressive CSI prediction. The prediction horizon depends on both of the CSI and geometry rather than a general IEEE 802.11 channel model with certain statistical channel-aging coefficients. Since Sionna ray tracing (RT) connects specified STA trajectories to multipath channel
realizations~\cite{Alkhateeb2019DeepMIMO,Hoydis2023SionnaRT,SionnaRTTechnical2025}, one can accurately verify the CSI prediction performance. In our framework, Sionna RT supplies the training and evaluation channels, while the learned predictor estimates
future CSI from historical CSI, with causal receiver-position information as an optional input. Then the prediction performance is evaluated through IEEE 802.11bn link-level simulation. Unlike prior work \cite{Zhang2026Sounding} with one-step single prediction on a single subcarrier, this paper takes a more ambitious approach to investigate the autoregressive multi-subcarrier CSI prediction in WiFi MIMO systems with mobile WiFi clients based on the compressed Wi-Fi beamforming feedback simulated through Sionna RT. The dynamic channel sounding interval is also optimized under signal-to-interference- plus-noise ratio (SINR) bound for quality of service (QoS) guarantee. The distinction to our earlier work \cite{Zhang2026Sounding} is therefore not only the channel generator and prediction manner, but also the explicit connection between channel-prediction error and feasible transmission over the selected sounding interval. Despite the advantages of the proposed channel sounding architecture, we have encountered multiple challenges. First, autoregressive CSI prediction is likely to introduce cumulative prediction error; therefore per-stream SINR must be formulated and controlled by the optimized channel sounding interval. Second, theoretically speaking, the autoregressive CSI prediction must combine the positional information and the CSI to achieve higher prediction performance. Third, the throughput objective must strike the balance between the CSI prediction performance and the prediction overhead. To this end, we make the following contributions.
\begin{enumerate}
    \item We develop an autoregressive CSI prediction framework
    combining ray-traced channels, temporal rotary position
    embedding (RoPE)~\cite{Su2021RoFormer}, and Huber-loss training.
    The predicted complex channel coefficients support beamforming
    updates between sounding exchanges, with optional spatial
    encoding from causal position information.

    \item We derive a per-stream SINR lower bound for
    prediction-based SVD precoding and combining. This bound
    yields sufficient feasibility conditions for sounding-interval
    selection, which maximizes predicted average throughput while
    accounting for sounding and processing overhead. We prove that cumulative-sum enumeration solves the sounding-interval problem, with linear search complexity after CSI prediction and SINR evaluation. The selected interval satisfies the actual SINR constraints whenever the channel errors remain within their assumed bounds.

    \item We evaluate channel prediction, transfer learning,
    autoregressive prediction error, and single-STA goodput with
    ground-truth, autoregressive prediction, and stale CSI. Predicted-CSI beamforming
    improves mean goodput and DL rate adaptation on all three evaluated trajectories,
    while error accumulation in autoregressive prediction and
    occasional performance reversals support a finite prediction
    horizon.
\end{enumerate}

The remainder of this paper is organized as follows.
Section~\ref{sec:system} presents the channel and sounding models.
Section~\ref{sec:proposed} develops the predictor, sounding-interval optimization,
and complexity analysis.
Section~\ref{sec:evaluation} evaluates CSI prediction and
link performance, and Section~\ref{sec:conclusion} concludes this paper.
Proofs are provided in the supplemental appendices.

\emph{Notation:}
Italic letters denote scalars, bold lowercase and uppercase
letters denote vectors and matrices, respectively, and bold
calligraphic letters denote channel tensors.
Other calligraphic symbols represent sets and quantities defined
locally.
The sets of real and complex numbers are denoted by
\(\mathbb{R}\) and \(\mathbb{C}\), respectively. We use
\(\R^{+}=[0,\infty)\), with
\(\mathrm{j}=\sqrt{-1}\).
Superscripts \((\cdot)^{\mathsf T}\),
\((\cdot)^{\mathsf H}\), and \((\cdot)^*\) denote transpose,
Hermitian transpose, and complex conjugation, respectively.
The operators \(\Re\{\cdot\}\) and \(\Im\{\cdot\}\) extract
the real and imaginary parts.
For a matrix \(\mathbf A\), \(\mathbf A(i,:)\),
\(\mathbf A(:,j)\), and \([\mathbf A]_{i,j}\) denote its
\(i\)th row, \(j\)th column, and \((i,j)\)th entry, respectively.
The matrix \(\mathbf I_d\) is the \(d\times d\) identity matrix.
The notation \(|\cdot|\) denotes scalar magnitude or set
cardinality, according to its argument.
The norm \(\|\cdot\|_{\mathrm F}\) denotes the Frobenius norm,
whereas \(\|\cdot\|_2\) denotes the Euclidean norm for vectors
and the spectral norm for matrices.
We write \(\mathbb{E}[\cdot]\) for expectation,
\([x]_+=\max\{x,0\}\), and \(\lceil x\rceil\) for the smallest
integer no less than \(x\).
The Kronecker delta \(\delta_{a,b}\) equals one when \(a=b\)
and zero otherwise. The indicator \(\mathbf1\{A\}\) equals one
when event \(A\) occurs and zero otherwise.
Hats denote predicted or estimated quantities, including trained model parameters.

\section{System Model}
\label{sec:system}
This section models MIMO transmission over time-varying channels
and quantifies the throughput cost of channel sounding and processing overhead.

\subsection{Time-Varying MIMO Channel}
An AP with $N_{\rm t}$ antennas communicates with
a mobile STA with $N_{\rm r}$ receive antennas over
$N_{\rm s}\in\mathbb Z^{++}$ spatial streams, where
$1\le N_{\rm s}\le\min\{N_{\rm t},N_{\rm r}\}$.
For each sounding cycle, we use the local time index $j$,
with $j=0$ at the start of sounding. Let $T_{\rm s}$ be the
slot duration and channel-sampling interval in seconds.
At time index $j$, let
$\mathbf r[j]\in\R^3$ denote the STA position and
$\mathbf H_q[j]\in\C^{N_{\rm r}\times N_{\rm t}}$ denote its MIMO channel
on subcarrier $q$. The indices $i=1,\ldots,N_{\rm s}$ and
$q=1,\ldots,N_{\rm sc}$ denote streams and subcarriers,
respectively. Let $h_q^{(u,v)}[j]\in\C$ be the coefficient from transmit antenna $v=1,\ldots,N_{\rm t}$ to receive antenna
$u=1,\ldots,N_{\rm r}$. The channel tensor $\boldsymbol{\mathcal H}[j]\in \C^{N_{\rm sc}\times N_{\rm r}\times N_{\rm t}}$.

\begin{figure}[!t]
\centering
\noindent\makebox[\linewidth][c]{%
\includegraphics[width=\linewidth]{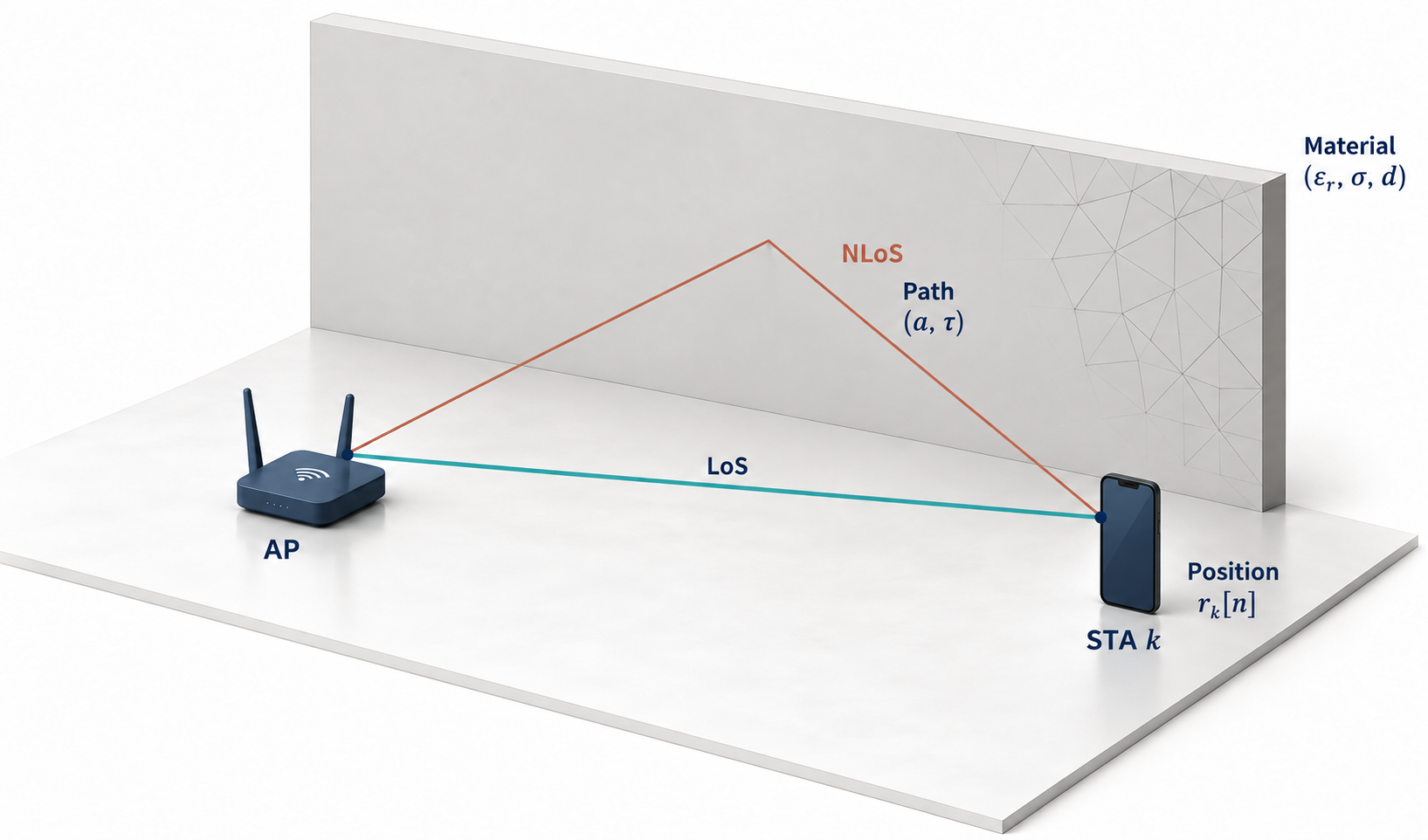}%
}\par
\caption{Ray-tracing geometry and the material, path, and position variables used in the channel model.}
\label{fig:channel_model}
\end{figure}

In Fig.~\ref{fig:channel_model}, $a\in\C$ is a path coefficient
and $\tau$ is its propagation delay in seconds. Reflecting
materials have relative permittivity $\varepsilon_r$, conductivity
$\sigma$, and thickness $d\ge0$ in meters. Let $\mathcal M$ denote the collection of these parameters, which determine the reflection response together with incidence angle and
polarization. Path geometry determines the propagation delay
\cite{Hoydis2023SionnaRT,SionnaRTTechnical2025}. 

\begin{figure*}[!t]
\centering
\noindent\makebox[\linewidth][c]{%
\includegraphics[width=0.88\linewidth,height=0.22\linewidth]{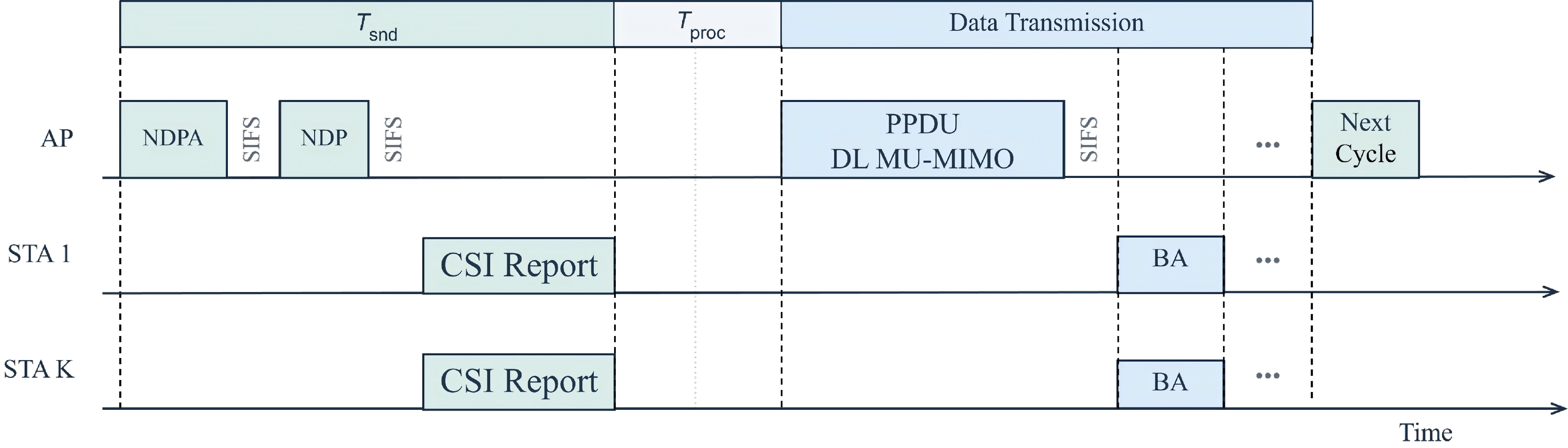}%
}\par
\caption{Frame exchange of channel sounding.}
\label{fig:sounding_timeline}
\end{figure*}

Let $\mathcal P[j]$ be the propagation-path set at local time
$jT_{\rm s}$. For $\ell\in\mathcal P[j]$ and antenna pair $(u,v)$,
let $\tau_{\ell}^{(u,v)}[j]\ge0$ be the geometric propagation
delay in seconds. Let $f_{\rm c}$ and $\Delta f>0$ be the carrier frequency and subcarrier spacing in hertz. Define $\nu_q=(q-\lfloor N_{\rm sc}/2\rfloor)\Delta f$ and
$f_q=f_{\rm c}+\nu_q$. When at least one valid path exists, let
$\tau_{\min}[j]$ be the earliest valid delay and define
$\bar\tau_{\ell}^{(u,v)}[j]
=\tau_{\ell}^{(u,v)}[j]-\tau_{\min}[j]\ge0$.
Define $a_{\ell}^{(u,v)}[j](\mathcal M)\in\C$ as the raw
passband path coefficient before applying the propagation-delay
phase. It includes propagation loss and material/polarization
response, but excludes the delay phase applied below. A coefficient
is zero if the path does not connect the antenna pair. Then
\begin{equation}
 \mathbf H_q[j]=[h_q^{(u,v)}[j]]_{u,v},
 \label{eq:mimo_matrix}
\end{equation}
where
\[
 h_q^{(u,v)}[j]=\sum_{\ell\in\mathcal P[j]}
 a_{\ell}^{(u,v)}[j](\mathcal M)
 e^{-\mathrm j2\pi(f_{\rm c}+\nu_q)
 \bar\tau_{\ell}^{(u,v)}[j]}.
\]
Equivalently, the baseband coefficient
$b_{\ell}^{(u,v)}[j]
=a_{\ell}^{(u,v)}[j](\mathcal M)
e^{-\mathrm j2\pi f_{\rm c}\bar\tau_{\ell}^{(u,v)}[j]}$
gives the same response by applying only
$e^{-\mathrm j2\pi\nu_q\bar\tau_{\ell}^{(u,v)}[j]}$ to $b$.
Thus, the coefficients that contain carrier-delay phase must
use the offset-frequency expression. The delay reference and
coefficient phase are transformed consistently. Channel energy is
not normalized~\cite{SionnaRTPaths2026}.

For individually traced antenna paths, the geometric antenna-pair
delays supply the spatial propagation phase. For synthetic arrays,
the alternative representation uses array-center path delays and
the corresponding steering responses; the same spatial phase cannot be introduced through individual antenna-pair delays.
Receive and transmit antennas index rows and columns, respectively. No rank-one assumption is imposed. If $\mathcal P[j]$ is empty, we set $\mathbf H_q[j]=\mathbf0$ without defining $\tau_{\min}[j]$.

Receiver motion changes path gains, delays, and visibility,
altering channel coefficients and spatial subspaces. Path appearance
or disappearance can cause abrupt changes. Define
$\Delta\mathbf H_q[j]=\mathbf H_q[j]-\mathbf H_q[0]$
as the variation from the current cycle's zero-time channel.
Prediction and error evaluation use the same channel phase
convention. No temporal correlation model is imposed.

\subsection{MIMO Transmission and Channel Sounding}
Let \(\mathbf V_q[j]\in\C^{N_{\rm t}\times N_{\rm s}}\) and
\(\mathbf W_q[j]\in\C^{N_{\rm s}\times N_{\rm r}}\) denote the
precoder and combiner at time index \(j\). The precoder
includes the stream powers, with
\(\mathbf V_q^{\mathsf H}[j]\mathbf V_q[j]
=\diag(p_{q,1}[j],\ldots,p_{q,N_{\rm s}}[j])\), where
\(p_{q,i}[j]>0\). Each combining row has a unit norm.
Let \(d_{q,i}[j]\in\C\) denote the \(i\)th entry of
\(\mathbf d_q[j]\in\C^{N_{\rm s}}\), transmitted on
subcarrier \(q\) at time index \(j\).

The data streams are zero-mean and uncorrelated, with unit average
symbol power: \(\E[d_{q,i}[j]]=0\) and
\(\E[d_{q,i}[j]d_{q,r}^*[j]]=\delta_{i,r}\),
where \(r=1,\ldots,N_{\rm s}\). The Kronecker delta
\(\delta_{a,b}\) equals one for \(a=b\) and zero for \(a\ne b\).
Expectations are taken over data and noise with the channel and
transceiver fixed.
Let \(\mathbf x_q[j]\in\C^{N_{\rm t}}\) denote the transmitted
signal vector, and let \(\mathbf y_q[j]\in\C^{N_{\rm r}}\)
denote the received signal vector at the STA.
The noise \(\mathbf n_q[j]\in\C^{N_{\rm r}}\) satisfies
\(\mathbf n_q[j]\perp\mathbf d_q[j]\),
\(\E[\mathbf n_q[j]]=\mathbf0\),
\(\E[\mathbf n_q[j]\mathbf n_q^{\mathsf H}[j]]
=\sigma_{\rm n}^2\mathbf I_{N_{\rm r}}\), where
\(\sigma_{\rm n}^2>0\) is the noise variance per receive antenna.
The transmitted and received signal vectors are
\begin{align}
 \mathbf x_q[j]
 &=\mathbf V_q[j]\mathbf d_q[j],
 \label{eq:transmit_signal}\\
 \mathbf y_q[j]
 &=\mathbf H_q[j]\mathbf x_q[j]
   +\mathbf n_q[j].
 \label{eq:mimo_received}
\end{align}

Let \(\widetilde{\mathbf y}_q[j]=\mathbf W_q[j]\mathbf y_q[j]
\in\C^{N_{\rm s}}\) denote the combined signal vector, and let
\(\widetilde{\mathbf n}_q[j]=\mathbf W_q[j]\mathbf n_q[j]
\in\C^{N_{\rm s}}\) denote the combined noise vector. Substituting \eqref{eq:transmit_signal} into
\eqref{eq:mimo_received} gives
\begin{equation}
 \widetilde{\mathbf y}_q[j]
 =\mathbf W_q[j]\mathbf H_q[j]\mathbf V_q[j]\mathbf d_q[j]
  +\widetilde{\mathbf n}_q[j].
 \label{eq:mimo_combined}
\end{equation}
Let \(P_q>0\) denote the transmit-power budget on subcarrier \(q\).
The unit-power, uncorrelated data streams give
\(\E\|\mathbf x_q[j]\|_2^2=\|\mathbf V_q[j]\|_{\rm F}^2
=\sum_{r=1}^{N_{\rm s}}p_{q,r}[j]\le P_q\). Let \(y_{q,i}[j]\) and \(\widetilde n_{q,i}[j]\in\C\) denote the
\(i\)th entries of \(\widetilde{\mathbf y}_q[j]\) and
\(\widetilde{\mathbf n}_q[j]\), respectively.
The \((i,r)\)th entry of \(\mathbf W_q[j]\mathbf H_q[j]\mathbf V_q[j]\)
gives the coupling from the transmitted stream \(r\) to received stream \(i\).
Expanding \eqref{eq:mimo_combined} yields
\begin{equation}
 \begin{aligned}
 &y_{q,i}[j]=[\mathbf W_q[j]\mathbf H_q[j]\mathbf V_q[j]]_{i,i}
                 d_{q,i}[j]\\
 &\quad+\sum_{\substack{r=1\\r\ne i}}^{N_{\rm s}}
 [\mathbf W_q[j]\mathbf H_q[j]\mathbf V_q[j]]_{i,r}d_{q,r}[j]
 +\widetilde n_{q,i}[j].
 \end{aligned}
 \label{eq:stream_signal}
\end{equation}
The terms in \eqref{eq:stream_signal} represent the desired signal, inter-stream interference,
and noise, respectively. For \(\mathbf H\in\C^{N_{\rm r}\times N_{\rm t}}\),
define the interference power
\(I_{q,i}[j](\mathbf H)=\allowbreak
\sum_{\substack{r=1\\r\ne i}}^{N_{\rm s}}
|[\mathbf W_q[j]\mathbf H\mathbf V_q[j]]_{i,r}|^2\in\R^{+}\).
Empty interference sums are zero. Unit-norm combining gives
\(\E[|\widetilde n_{q,i}[j]|^2]=\sigma_{\rm n}^2\).
The per-stream SINR
\(\gamma_{q,i}[j](\mathbf H)\in\R^{+}\) is
\begin{equation}
 \gamma_{q,i}[j](\mathbf H)=
 \frac{|[\mathbf W_q[j]\mathbf H\mathbf V_q[j]]_{i,i}|^2}
 {\sigma_{\rm n}^2+I_{q,i}[j](\mathbf H)}.
 \label{eq:sinr}
\end{equation}

The frame exchange of channel sounding in Fig.~\ref{fig:sounding_timeline} acquires
CSI through training and feedback~\cite{Zhang2026Sounding}.
Let \(T_{\rm snd}>0\) denote its total airtime in seconds,
including frame
overhead and interframe spaces. Let
\(\widetilde{\mathbf H}_q[0]
=\mathbf H_q[0]+\mathbf E^{\rm obs}_q[0]
\in\C^{N_{\rm r}\times N_{\rm t}}\) denote the sounded estimate
referenced to the cycle-start channel. The estimate is available
after channel sounding. The error \(\mathbf E^{\rm obs}_q[0]\) includes
CSI estimation, reporting, and acquisition-to-reference mismatch.
This reference alignment is a modeling assumption.
Let \(T_{\rm proc}\ge0\) denote processing time in seconds.
The cycle contains slots \(t=1,\ldots,T\), with slot \(t\)
starting at local channel index \(j=t-1\).
Slots \(1,\ldots,T_0-1\) are reserved for sounding and processing and
slots \(T_0,\ldots,T\) are used for carrying data.
Thus the first data sample has local index \(j=T_0-1\).
The sounding interval \(T\) is the only optimization variable.
Appendix~A gives the overhead accounting.
\begin{figure*}[!t]
\centering
\noindent\makebox[\linewidth][c]{%
\includegraphics[width=\textwidth]{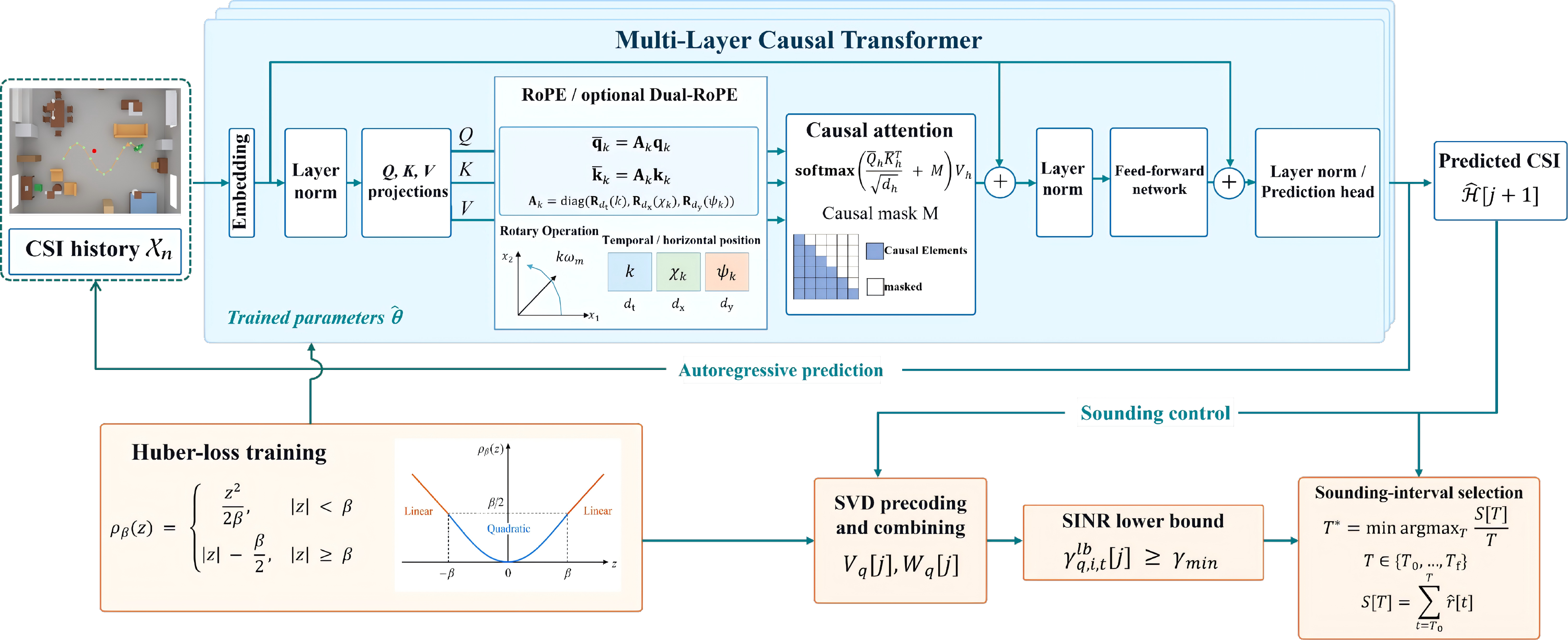}%
}\par
\caption{Autoregressive CSI prediction, Huber-loss training, and prediction-aided sounding control.}
\label{fig:proposed_architecture}
\end{figure*}

Let \(\mathbf V_q^0\) and \(\mathbf W_q^0\)
denote the precoding and combining matrices computed from the
observed CSI \(\widetilde{\mathbf H}_q[0]\) and held fixed until
the next sounding.
The channel mismatch is
\(\mathbf D_q[j]=\mathbf H_q[j]-
\widetilde{\mathbf H}_q[0]
=\Delta\mathbf H_q[j]-\mathbf E^{\rm obs}_q[0]
\in\C^{N_{\rm r}\times N_{\rm t}}\).
Define \(I_{q,i}^0(\mathbf H)\) as the interference by replacing
\(\mathbf W_q[j]\) and \(\mathbf V_q[j]\) in
\(I_{q,i}[j](\mathbf H)\) with
\(\mathbf W_q^0\) and \(\mathbf V_q^0\),
respectively. For
\(\mathbf H=\widetilde{\mathbf H}_q[0]+\mathbf D_q[j]\),
\eqref{eq:sinr} gives
\begin{equation}
 \gamma_{q,i}^{\rm stale}[j]=
 \frac{|[\mathbf W_q^0\mathbf H\mathbf V_q^0]_{i,i}|^2}
 {\sigma_{\rm n}^2+I_{q,i}^0(\mathbf H)}.
 \label{eq:aging_sinr}
\end{equation}
The mismatch adds
\([\mathbf W_q^0\mathbf D_q[j]\mathbf V_q^0]_{i,r}\)
to the coupling from stream \(r\) to stream \(i\). It can reduce
the desired gain and introduce inter-stream interference, lowering
SINR and goodput. More frequent sounding limits this mismatch but
leaves less airtime for data transmission.

\subsection{Throughput}
For data slot \(t\), let \(\mathbf H_{q,t}=\mathbf H_q[t-1]\in
\C^{N_{\rm r}\times N_{\rm t}}\) denote the channel matrix on subcarrier
\(q\), and \(\gamma_{q,i,t}(\mathcal H)\in\R^{+}\)
denote the SINR of stream \(i\), obtained from \eqref{eq:sinr}
using this slot's channel, precoder, and combiner. Here
\(\mathcal H=\{\mathbf H_{q,t}:0\le q<N_{\rm sc},\,
T_0\le t\le T_{\max}\}\) is the channel sequence over the
prediction horizon. Slot \(t\) corresponds to predictor sample
\(j=t-1\); the sampling index is not a control variable.
Let \(c_{\rm r}>0\) denote
the fixed data-symbol rate including OFDM and packet overhead.
Let \(r[t](\mathcal H)\in\R^{+}\) denote the throughput
in slot \(t\), and let \(R(T;\mathcal H)\in\R^{+}\)
denote the average throughput over a sounding cycle, both in bit/s.
They are given by
\begin{align}
 r[t](\mathcal H)
 &=c_{\rm r}\sum_{q=0}^{N_{\rm sc}-1}\sum_{i=1}^{N_{\rm s}}
   \log_2\bigl(1+\gamma_{q,i,t}(\mathcal H)\bigr),
 \label{eq:data_rate}\\
 R(T;\mathcal H)
 &=\frac{1}{T}\sum_{t=T_0}^{T}r[t](\mathcal H).
 \label{eq:utility}
\end{align}
The denominator \(T\) includes all cycle slots; the overhead slots
contribute no transmitted data. The common slot duration therefore
cancels when transmitted data are divided by cycle duration.
These SINR-based throughput expressions do not include packet decoding
or retransmissions.

\section{Prediction-aided Sounding Framework}
\label{sec:proposed}\label{sec:predictor}\label{sec:control}
This section develops joint CSI prediction across subcarriers, formulates SINR-constrained sounding-interval optimization, and derives a globally optimal
algorithm for the resulting finite problem.

\subsection{Autoregressive CSI Prediction}
\label{sec:learning}
The information \(\mathcal X\), available after sounding, includes the
\(L\)-snapshot channel history
\(\mathcal B=(\boldsymbol{\mathcal B}_{0},\ldots,
\boldsymbol{\mathcal B}_{L-1})\),
where \(\boldsymbol{\mathcal B}_{\ell}\in
\C^{N_{\rm sc}\times N_{\rm r}\times N_{\rm t}}\)
denotes the available CSI at time index \(\ell=0,\ldots,L-1\). Unobserved entries contain predictions,
and the latest entry is the observed tensor
\(\widetilde{\boldsymbol{\mathcal H}}[0]
=\{\widetilde{\mathbf H}_q[0]\}_q\).
Let
\(\widehat{\boldsymbol{\mathcal H}}[j]
=\{\widehat{\mathbf H}_q[j]\}_q\)
collect the per-subcarrier predictions conditioned on \(\mathcal X\) into a complex tensor in
\(\C^{N_{\rm sc}\times N_{\rm r}\times N_{\rm t}}\).
The reference prediction is
\(\widehat{\boldsymbol{\mathcal H}}[0]
=\widetilde{\boldsymbol{\mathcal H}}[0]\).
Let \(\operatorname{tok}:\C^{N_{\rm sc}\times N_{\rm r}\times N_{\rm t}}
\to\R^{D_0}\) denote the invertible map that concatenates the
real channel coefficients followed by the imaginary coefficients, where \(D_0=2N_{\rm sc}N_{\rm r}N_{\rm t}\).
Each token represents all selected subcarriers at one time step. The history CSI matrix \(\mathbf X\in\R^{L\times D_0}\) has entries
\([\mathbf X]_{\ell,c}
=[\operatorname{tok}(\boldsymbol{\mathcal B}_{\ell})]_c\),
for \(\ell=0,\ldots,L-1\) and \(c=0,\ldots,D_0-1\).
More generally, \(\mathbf X\in\R^{L_{\rm in}\times D_0}\)
contains \(L_{\rm in}\) tokens, with row index
\(k=0,\ldots,L_{\rm in}-1\).
For fixed real offsets \(\mu_c\in\R\) and positive real scales
\(\lambda_c>0\),
\(c=0,\ldots,D_0-1\), define feature-wise normalization by
\([\mathcal N(\mathbf X)]_{k,c}=([\mathbf X]_{k,c}-\mu_c)/\lambda_c\)
and its inverse by
\([\mathcal N^{-1}(\mathbf X)]_{k,c}=\lambda_c[\mathbf X]_{k,c}+\mu_c\).
Let \(\boldsymbol\theta\) denote the real parameter vector of
the causal Transformer. The predictor \(\mathcal F_{\boldsymbol\theta}\)
applies \(\mathcal N\), the Transformer, and \(\mathcal N^{-1}\),
so its output is expressed in channel units.

Let \(\mathbf P\in\R^{L_{\rm in}\times3}\) denote the STA
position. The spatial
predictor returns
\(\mathcal F_{\boldsymbol\theta}(\mathbf X;\mathbf P)
\in\R^{L_{\rm in}\times D_0}\); the temporal predictor omits
\(\mathbf P\). Let \(\mathbf F^{(\ell)}\in\R^{L_{\rm in}\times d}\)
denote the features at layer \(\ell=0,\ldots,N_\ell\).
As presented in Fig. \ref{fig:proposed_architecture}, the network has an embedding dimension \(d\), \(N_\ell\) decoder
layers, \(N_{\rm h}\) attention heads of even dimension
\(d_{\rm h}=d/N_{\rm h}\), and a feed-forward network (FFN) of
width \(d_{\rm ff}\) with Gaussian error linear unit (GELU)
activation~\cite{Vaswani2017}. The head and token indices are
\(h=1,\ldots,N_{\rm h}\) and \(k,k'=0,\ldots,L_{\rm in}-1\).
Embedding and reconstruction are
\begin{align}
 \mathbf F^{(0)}
 &=\mathcal N(\mathbf X)\mathbf W_{\rm in}
    +\mathbf1\mathbf b_{\rm in}^{\mathsf T},
 \label{eq:input_embedding}\\
 \mathcal F_{\boldsymbol\theta}(\mathbf X;\mathbf P)
 &=\mathcal N^{-1}\!\left(
 \operatorname{LN}(\mathbf F^{(N_\ell)})\mathbf W_{\rm out}
 +\mathbf1\mathbf b_{\rm out}^{\mathsf T}\right),
 \label{eq:predictor_io}
\end{align}
where \(\mathbf W_{\rm in}\in\R^{D_0\times d}\) and \(\mathbf W_{\rm out}\in\R^{d\times D_0}\) are learned projection matrices,
\(\mathbf b_{\rm in}\in\R^d\) and \(\mathbf b_{\rm out}\in\R^{D_0}\)
are learned biases, and \(\mathbf1\)
is the all-ones vector in \(\R^{L_{\rm in}}\).
\(\operatorname{LN}\) denotes layer normalization across the
features of each row. The embedding combines the frequency and
MIMO channel coordinates; causal attention operates over time.

For a positive even integer \(d'\) and \(z\in\R\), let
\(\mathbf R_{d'}(z)\in\R^{d'\times d'}\) denote the block-diagonal
matrix of two-dimensional rotations with angles \(z\omega_m\),
where \(\omega_m>0\), \(m=0,\ldots,d'/2-1\), are fixed
rotary frequencies.
Temporal rotary position embedding (RoPE)~\cite{Su2021RoFormer}
uses the rotation matrix
\(\mathbf A_k=\mathbf R_{d_{\rm h}}(k)\in\R^{d_{\rm h}\times d_{\rm h}}\)
at token \(k\).
At layer \(\ell=1,\ldots,N_\ell\), let
\(\mathbf F=\mathbf F^{(\ell-1)}\). Head \(h\) affinely
projects \(\operatorname{LN}(\mathbf F)\) into queries, keys, and
values \(\mathbf Q_h,\mathbf K_h,\mathbf V_h\in\R^{L_{\rm in}\times d_{\rm h}}\), respectively. Let \(\mathbf q_k,\mathbf k_k\in\R^{d_{\rm h}}\) denote the
transposed \(k\)th rows of \(\mathbf Q_h\) and \(\mathbf K_h\),
respectively;
the encoding rotates
\(\overline{\mathbf q}_k=\mathbf A_k\mathbf q_k\) and
\(\overline{\mathbf k}_k=\mathbf A_k\mathbf k_k\), leaving values
unchanged. The identity
\(\mathbf R_{d'}(z)^{\mathsf T}\mathbf R_{d'}(z')
=\mathbf R_{d'}(z'-z)\) gives
\begin{equation}
 \overline{\mathbf q}_k^{\mathsf T}\overline{\mathbf k}_{k'}
 =\mathbf q_k^{\mathsf T}
   \mathbf R_{d_{\rm h}}(k'-k)\mathbf k_{k'}.
 \label{eq:relative_rotation}
\end{equation}
The RoPE block in Fig.~\ref{fig:proposed_architecture} encodes
relative sample displacement through the rotated query--key product
in \eqref{eq:relative_rotation}. Dual-RoPE partitions each head into temporal and horizontal spatial
components of positive even dimensions \(d_{\rm t},d_{\rm x},d_{\rm y}\), with
\(d_{\rm t}+d_{\rm x}+d_{\rm y}=d_{\rm h}\). For the spatial encoding, let \((x_k,y_k)\) be the first two coordinates of
the \(k\)th row of \(\mathbf P\). A fixed reference origin
\((x_{\rm ref},y_{\rm ref})\in\R^2\) and length
scale \(\ell_{\rm ref}>0\), both in meters, define the
dimensionless real coordinates
\(\chi_k=(x_k-x_{\rm ref})/\ell_{\rm ref}\) and
\(\psi_k=(y_k-y_{\rm ref})/\ell_{\rm ref}\), giving
\(\mathbf A_k=\operatorname{diag}\!\bigl(
\mathbf R_{d_{\rm t}}(k),\mathbf R_{d_{\rm x}}(\chi_k),
\mathbf R_{d_{\rm y}}(\psi_k)\bigr)\).
Let
\(\mathbf q_{k,\alpha},\mathbf k_{k,\alpha}\in\R^{d_\alpha}\)
denote the query and key components in block
\(\alpha\in\{\mathrm{t},\mathrm{x},\mathrm{y}\}\). Define the displacements
\(\Delta_{\rm t}=k'-k\),
\(\Delta_{\rm x}=\chi_{k'}-\chi_k\),
and \(\Delta_{\rm y}=\psi_{k'}-\psi_k\).
The blockwise rotation identity yields
\begin{equation}
 \overline{\mathbf q}_k^{\mathsf T}\overline{\mathbf k}_{k'}
 =\sum_{\alpha\in\{\mathrm{t},\mathrm{x},\mathrm{y}\}}
 \mathbf q_{k,\alpha}^{\mathsf T}
 \mathbf R_{d_\alpha}(\Delta_\alpha)\mathbf k_{k',\alpha}.
 \label{eq:dual_score}
\end{equation}
For fixed queries and keys, these scores depend on relative
temporal and spatial displacement and are invariant to common
coordinate shifts. The rotations preserve query and key norms.
For either encoding, the rotated query and key matrices
\(\overline{\mathbf Q}_h,\overline{\mathbf K}_h
\in\R^{L_{\rm in}\times d_{\rm h}}\) are obtained by stacking
the rotated vectors as rows:
\begin{align*}
 \overline{\mathbf Q}_h
 &=\bigl[\mathbf A_0\mathbf q_0,\ldots,
         \mathbf A_{L_{\rm in}-1}\mathbf q_{L_{\rm in}-1}\bigr]^{\mathsf T},\\
 \overline{\mathbf K}_h
 &=\bigl[\mathbf A_0\mathbf k_0,\ldots,
         \mathbf A_{L_{\rm in}-1}\mathbf k_{L_{\rm in}-1}\bigr]^{\mathsf T}.
\end{align*}
Thus \([\overline{\mathbf Q}_h\overline{\mathbf K}_h^{\mathsf T}]_{k,k'}
=\overline{\mathbf q}_k^{\mathsf T}\overline{\mathbf k}_{k'}\).
The attention-head output
\(\mathbf O_h\in\R^{L_{\rm in}\times d_{\rm h}}\) is
\begin{equation}
 \mathbf O_h=\operatorname{softmax}\!\left(
 \frac{\overline{\mathbf Q}_h\overline{\mathbf K}_h^{\mathsf T}}
      {\sqrt{d_{\rm h}}}+\mathbf M\right)\mathbf V_h,
 \label{eq:attention}
\end{equation}
where softmax is applied rowwise. The causal mask
\(\mathbf M\in\{0,-\infty\}^{L_{\rm in}\times L_{\rm in}}\) satisfies \([\mathbf M]_{k,k'}=0\) for
\(k'\le k\), and \([\mathbf M]_{k,k'}=-\infty\) otherwise.
The attention output \(\mathbf O\in\R^{L_{\rm in}\times d}\)
is the projected concatenation
\(\mathbf O=[\mathbf O_1|\cdots|\mathbf O_{N_{\rm h}}]\mathbf W_{\rm O}\),
where \(\mathbf W_{\rm O}\in\R^{d\times d}\) is the learned output
projection and the brackets concatenate the attention-head outputs.
At inference, the intermediate and output feature matrices
\(\mathbf F',\mathbf F^{\rm out}\in\R^{L_{\rm in}\times d}\) are
\begin{align}
 \mathbf F'&=\mathbf F+\mathbf O,
 \label{eq:attention_residual}\\
 \mathbf F^{\rm out}&=\mathbf F'
             +\operatorname{FFN}(\operatorname{LN}(\mathbf F')).
 \label{eq:residual_updates}
\end{align}
The layer output is \(\mathbf F^{\rm out}=\mathbf F^{(\ell)}\),
\(\ell=1,\ldots,N_\ell\).
The rowwise FFN applies learned affine maps
\(\R^d\to\R^{d_{\rm ff}}\to\R^d\), with GELU between them.
The rowwise operations and mask restrict output row \(k\)
to inputs \(0,\ldots,k\).

For training window \(w\), let
\(\mathbf X_w\in\R^{L_{\rm in}\times D_0}\) contain \(L\)
historical joint-channel tokens followed by the first
\(N_{\rm tar}-1\) target tokens, where
\(L_{\rm in}=L+N_{\rm tar}-1\). Let
\(\mathbf P_w\in\R^{L_{\rm in}\times3}\) contain the paired
STA positions for the spatial variant. The predicted-token matrix is
\(\widehat{\mathbf Y}_w
=\mathcal F_{\boldsymbol\theta}(\mathbf X_w;\mathbf P_w)
\in\R^{L_{\rm in}\times D_0}\), with \(\mathbf P_w\) omitted
for temporal RoPE. Let \(\mathbf y_{w,r}\in\R^{D_0}\) denote
target \(r\) in window \(w\). Its prediction satisfies
\(\widehat y_{w,r,c}=[\widehat{\mathbf Y}_w]_{L+r-2,c}\),
for \(r=1,\ldots,N_{\rm tar}\) and \(c=0,\ldots,D_0-1\).
Causality restricts the prediction of target \(r\) to the history and its
\(r-1\) preceding targets. For \(M_{\rm tr}\) training windows,
define \(z_{w,r,c}=y_{w,r,c}-\widehat y_{w,r,c}\in\R\).
The Huber-loss branch in Fig.~\ref{fig:proposed_architecture}
uses the empirical objective \(\mathcal L(\boldsymbol\theta)\in\R^{+}\):
\begin{equation}
 \mathcal L(\boldsymbol\theta)=
 \frac{\displaystyle\sum_{w=1}^{M_{\rm tr}}
       \sum_{r=1}^{N_{\rm tar}}\sum_{c=0}^{D_0-1}
       \rho_\beta(z_{w,r,c})}
      {M_{\rm tr}N_{\rm tar}D_0}.
 \label{eq:training_problem}
\end{equation}
The scaled
Huber loss weight \(\rho_\beta:\R\to\R^{+}\)~\cite{Huber1964} is \(\rho_\beta(z)=z^2/(2\beta)\)
when \(|z|<\beta\) and \(|z|-\beta/2\) otherwise, where $\beta > 0$. Mean square error (MSE) variants
use \(\rho(z)=z^2\). The Huber derivative
\(\rho_\beta'(z)=\operatorname{sgn}(z)\min\{|z|/\beta,1\}\)
is bounded in magnitude by one. Let \(\widehat{\boldsymbol\theta}\) denote
the trained parameter vector. This does not imply a deterministic
bound on autoregressive prediction error. The objective is nonconvex in
\(\boldsymbol\theta\).

Between soundings, the predictor appends each CSI prediction
to the \(L\)-snapshot history. Let \(\mathcal B^{(0)}=\mathcal B\) and set \(L_{\rm in}=L\).
For the spatial predictor, \(\mathbf P^{(j)}\in\R^{L\times3}\)
contains the paired position history, initialized with the measured
or previously predicted positions in \(\mathcal X\).
Let \(f_{\widehat{\boldsymbol\theta}}\) apply
\(\mathcal F_{\widehat{\boldsymbol\theta}}\) once to the tokenized
joint-channel history and reconstruct the last output with the inverse mapping \(\operatorname{tok}^{-1}\). The resulting channel tensor is
\begin{equation}
 \widehat{\boldsymbol{\mathcal H}}[j+1]
   =f_{\widehat{\boldsymbol\theta}}(\mathcal B^{(j)};\mathbf P^{(j)}) \in \C^{N_{\rm sc}\times N_{\rm r}\times N_{\rm t}},
 \label{eq:block_prediction}
\end{equation}
with \(\mathbf P^{(j)}\) omitted for temporal RoPE. The feedback path in Fig.~\ref{fig:proposed_architecture}
appends this tensor to the history and discards its oldest entry,
yielding \(\mathcal B^{(j+1)}\). Each step advances by
\(T_{\rm s}\), producing the sequence used by all candidate intervals.
The rollout includes the overhead period and the first data forecast
is \(\widehat{\boldsymbol{\mathcal H}}[T_0-1]\).

For the causal spatial extension, let
\(\widetilde{\mathbf r}_1,\widetilde{\mathbf r}_2\in\R^3\) denote the latest two measured STA positions, acquired at
local time indices \(t_1<t_2=0\), and available before
controller processing. The estimated velocity
vector, in meters per second, is
\(\widehat{\mathbf v}=(\widetilde{\mathbf r}_2-
\widetilde{\mathbf r}_1)/(t_2-t_1)\in\R^3\).
The position prediction
\(\widehat{\mathbf r}[j+1]=\widetilde{\mathbf r}_2+
(j+1)T_{\rm s}\widehat{\mathbf v}\in\R^3\) is appended to the
position history, whose oldest entry is discarded, yielding
\(\mathbf P^{(j+1)}\) alongside the channel-history update.
The error analysis uses the same position extrapolation,
autoregressive history updates, and observation-error model.

\subsection{SINR Analysis and Problem Formulation}
\label{sec:error_bounds}\label{sec:problem}
The predicted channels determine the precoders and combiners. The prediction error then changes the desired gain and introduces inter-stream interference. We derive a SINR lower bound to construct
the sounding-interval constraints. The SVD of each predicted channel is
\begin{equation}
 \widehat{\mathbf H}_q[j]
 =\mathbf L_q[j]\widehat{\boldsymbol\Sigma}_q[j]\mathbf R_q^{\mathsf H}[j].
 \label{eq:svd}
\end{equation}
Here the complex matrices \(\mathbf L_q[j]\) and \(\mathbf R_q[j]\) have
\(N_{\rm r}\) and \(N_{\rm t}\) rows, respectively, and
\(\min\{N_{\rm r},N_{\rm t}\}\) orthonormal columns contain the left and right singular vectors, respectively.
The real diagonal matrix \(\widehat{\boldsymbol\Sigma}_q[j]\)
contains the singular values
\(\widehat s_{q,1}[j]\ge\cdots\ge
\widehat s_{q,\min\{N_{\rm r},N_{\rm t}\}}[j]\ge0\).
For \(i=1,\ldots,N_{\rm s}\), the precoder and combiner are
defined by
\([\mathbf V_q[j]]_{v,i}
=\sqrt{p_{q,i}[j]}[\mathbf R_q[j]]_{v,i}\) and
\([\mathbf W_q[j]]_{i,u}=[\mathbf L_q[j]]_{u,i}^{*}\).
On the predicted channel,
\(\gamma_{q,i}[j](\widehat{\mathbf H}_q[j])
=p_{q,i}[j]\widehat s_{q,i}^{\,2}[j]/\sigma_{\rm n}^2\).
The actual channel may retain inter-stream interference and is
evaluated using \eqref{eq:sinr}.

For subcarrier \(q\) and sample \(j\), let
\(\mathbf E_q[j]=\mathbf H_q[j]
-\widehat{\mathbf H}_q[j]\in\C^{N_{\rm r}\times N_{\rm t}}\)
denote the prediction error.
At \(j=0\), \(\mathbf E_q[0]=-\mathbf E^{\rm obs}_q[0]\).
Let \(\mathbf W=\mathbf W_q[j]\), \(\mathbf V=\mathbf V_q[j]\),
\(\mathbf E=\mathbf E_q[j]\), and \(p_r=p_{q,r}[j]\) for fixed $q$ and $j$.
Let \(g_i=\sqrt{p_i}\widehat s_{q,i}[j]\in\R^{+}\) denote
the predicted desired-signal amplitude and let
\(e_{i,r}=[\mathbf W\mathbf E\mathbf V]_{i,r}\in\C\) denote
the error in the coupling from stream \(r\) to stream \(i\).
The SVD diagonalizes the predicted channel, so the actual SINR
\(\gamma_i=\gamma_{q,i}[j](\mathbf H_q[j])\) is
\begin{equation}
 \gamma_i=\frac{|g_i+e_{i,i}|^2}
 {\sigma_{\rm n}^2+\sum_{\substack{r=1\\r\ne i}}^{N_{\rm s}}|e_{i,r}|^2}.
 \label{eq:su_sinr}
\end{equation}
Let \(\varepsilon_q[j]\in\R^{+}\) denote a given error bound,
\(\|\mathbf E_q[j]\|_{\rm F}\le\varepsilon_q[j]\), abbreviated
by \(\varepsilon\) for the fixed $q$ and $j$.
These bounds are fixed across candidate sounding intervals.
\begin{theorem}
\label{thm:sinr}
If \(\|\mathbf E\|_{\rm F}\le\varepsilon\), the SVD
precoder and combiner satisfy
\begin{equation}
 \gamma_i\ge\gamma_i^{\rm lb}(\varepsilon)
 :=\frac{[g_i-\varepsilon\sqrt{p_i}]_+^2}
 {\sigma_{\rm n}^2+
 \varepsilon^2\sum_{\substack{r=1\\r\ne i}}^{N_{\rm s}}p_r}.
 \label{eq:sinr_bound}
\end{equation}
\begin{proof}
The proof is shown in Appendix~B.
\end{proof}
\end{theorem}
The bound accounts for both desired-signal loss and inter-stream
interference. It provides a sufficient SINR condition and failure
of the condition does not necessarily imply link failure on the actual channel.

Let \(\widehat{\mathbf H}_{q,t}
=\widehat{\mathbf H}_q[t-1]\) and
\(\widehat{\mathcal H}=\{\widehat{\mathbf H}_{q,t}:0\le q<N_{\rm sc},\,
T_0\le t\le T_{\max}\}\) denote the predicted channel and
sequence in cycle-slot notation. Let
\(\widehat\gamma_{q,i,t}
=\gamma_{q,i,t}(\widehat{\mathcal H})\in\R^{+}\) denote the
predicted SINR, and \(\widehat r[t]=r[t](\widehat{\mathcal H})\in\R^{+}\) denote
the predicted slot throughput in bit/s. The predicted average
throughput is \(\widehat R(T)=R(T;\widehat{\mathcal H})\).
For data slot \(t\), let \(\gamma_{q,i,t}^{\rm lb}\in\R^{+}\)
represent the lower bound on the SINR evaluated at \(j=t-1\).
With \(\gamma_{\min}>0\), the sounding-interval selection problem is formulated as
\begin{subequations}\label{eq:robust_problem}
\begin{align}
 (\mathcal P_1)\quad&\max_{T\in\{T_0,\ldots,T_{\max}\}}
 \ \frac{1}{T}\sum_{t=T_0}^{T}\widehat r[t]
 \label{eq:robust_objective}\\
 \text{s.t.}\quad&\gamma_{q,i,t}^{\rm lb}\ge\gamma_{\min},
 \quad\forall q,i,\ t=T_0,\ldots,T.
 \label{eq:robust_guard}
\end{align}
\end{subequations}
The objective is the predicted average throughput in
\eqref{eq:utility}; the constraints apply to
\(q=0,\ldots,N_{\rm sc}-1\) and \(i=1,\ldots,N_{\rm s}\).
By Theorem~\ref{thm:sinr}, the constraints ensure actual SINR
feasibility whenever the channel errors satisfy their bounds.
If \(\rank(\widehat{\mathbf H}_{q,t})<N_{\rm s}\), a selected
singular value and its SINR lower bound are zero, violating
\eqref{eq:robust_guard}. Hence, no separate predicted-channel rank
constraint is required.

\subsection{Sounding-Interval Optimization}
\label{sec:interval_solution}
A failed SINR constraint excludes the possibility of including any further slot into the sounding interval. Thus, the feasible set of \(\mathcal P_1\)
is empty or \(\{T_0,\ldots,T_{\rm f}\}\), where \(T_{\rm f}\) is the
last feasible endpoint when it exists. Let \(S[T]=\sum_{t=T_0}^{T}\widehat r[t]\), with
\(S[T_0-1]=0\), since slots \(1,\ldots,T_0-1\) are reserved for channel sounding and processing. The optimization objective can be expressed as the average data rate \(S[T]/T\). Consider the data rate degradation constraint
\begin{equation}
 \widehat r[t+1]\le\widehat r[t].
 \label{eq:rate_degradation}
\end{equation}
A sufficient condition for \eqref{eq:rate_degradation} can
be expressed directly using the predicted SINRs already
computed from the channel, precoding, and combining matrices.
Specifically, evaluating \eqref{eq:sinr} on the predicted
channel gives
\begin{equation*}
 \widehat\gamma_{q,i,t}
 =
 \frac{
 |[\mathbf W_q[j]\widehat{\mathbf H}_{q,t}
       \mathbf V_q[j]]_{i,i}|^2
 }{
 \sigma_{\rm n}^2+
 I_{q,i}[j](\widehat{\mathbf H}_{q,t})
 }.
\end{equation*}
If these predicted SINRs satisfy $\widehat\gamma_{q,i,t+1}\le\widehat\gamma_{q,i,t}$, $\forall q,i$ at every consecutive pair of feasible data slots, then
\eqref{eq:data_rate} yields
\begin{align}
 \widehat r[t+1]-\widehat r[t]
 &=
 c_{\rm r}
 \sum_{q=0}^{N_{\rm sc}-1}
 \sum_{i=1}^{N_{\rm s}}
 \log_2\!\left(
 \frac{1+\widehat\gamma_{q,i,t+1}}
      {1+\widehat\gamma_{q,i,t}}
 \right) \le0.
\end{align}
The inequality holds because each ratio is at most one,
so every logarithm is non-positive, and \(c_{\rm r}>0\).
Thus, non-increasing predicted SINRs for all streams and
subcarriers are sufficient for \eqref{eq:rate_degradation}.
This condition is not necessary, since increases in some
stream rates may be outweighed by decreases in others. Algorithm~\ref{alg:control} checks \eqref{eq:rate_degradation}
on the feasible prediction sequence before binary search. The throughput increment can be defined as
\begin{equation}
 G(T)=T\widehat r[T+1]-S[T],\qquad T_0\le T<T_{\rm f}.
 \label{eq:interval_search_function}
\end{equation}

\begin{theorem}
\label{thm:selection}
Algorithm~\ref{alg:control} returns the smallest globally optimal
solution of \(\mathcal P_1\) in both branches, and returns \(0\)
if the feasible set is empty.
For a non-empty feasible set satisfying \eqref{eq:rate_degradation},
\(\widehat R(T)\) is unimodal. Its smallest global maximizer is
the first \(T<T_{\rm f}\) with \(G(T)\le0\), or
\(T_{\rm f}\) if no such endpoint exists. Any non-zero output satisfies the
actual SINR constraints whenever
\(\|\mathbf E_q[j]\|_{\rm F}\le\varepsilon_q[j]\)
for all subcarriers and prediction samples.
\begin{proof}
The proof is shown in Appendix~C.
\end{proof}
\end{theorem}

\begin{algorithm}[!t]
\small
\DontPrintSemicolon
\caption{Global Sounding-Interval Optimization}
\label{alg:control}
\KwIn{$\widehat{\mathcal H},\,\{p_{q,i}[j]\},\,
       \{\varepsilon_q[j]\},\,T_0,T_{\max},\gamma_{\min}$}
\KwOut{$T^*$}
$T_{\rm f}\leftarrow T_0-1$; $S[T_0-1]\leftarrow0$\;
\For{$t=T_0,\ldots,T_{\max}$}{
 Evaluate \eqref{eq:svd}, \eqref{eq:sinr_bound}\;
 \lIf{$\min_{q,i}\gamma_{q,i,t}^{\rm lb}<\gamma_{\min}$}{\textbf{break}}
 $\widehat r[t]\leftarrow r[t](\widehat{\mathcal H})$\;
 $S[t]\leftarrow S[t-1]+\widehat r[t]$; $T_{\rm f}\leftarrow t$\;
}
\lIf{$T_{\rm f}<T_0$}{\Return $0$}
\If{\eqref{eq:rate_degradation} fails}{
 \Return $\displaystyle
 \min\argmax_{t\in\{T_0,\ldots,T_{\rm f}\}}\frac{S[t]}{t}$\;
}
$a\leftarrow T_0$; $b\leftarrow T_{\rm f}$\;
\While{$a<b$}{
 $m\leftarrow\lfloor(a+b)/2\rfloor$\;
 \lIf{$G(m)\le0$}{$b\leftarrow m$}
 \lElse{$a\leftarrow m+1$}
}
\Return $a$\;
\end{algorithm}

Algorithm~\ref{alg:control} uses binary search when
\eqref{eq:rate_degradation} holds. For unrestricted prediction
sequences, its alternative branch compares all feasible averages. The proposed framework uses autoregressive CSI predictions to select a sounding interval that maximizes predicted average
throughput while accounting for sounding and processing overhead.
SINR lower bounds convert prediction-error bounds into sufficient
feasibility constraints, and Algorithm~\ref{alg:control} globally
solves \(\mathcal P_1\).

\begin{remark}
For an \(L\)-token history, one prediction costs
\(C_{\rm step}=O(LD_0d+N_\ell[Ld^2+L^2d+Ldd_{\rm ff}])\). Per-slot SVD costs \(O(N_{\rm sc}N_{\rm r}N_{\rm t}
\min\{N_{\rm r},N_{\rm t}\})\), and SINR-bound evaluation costs
\(O(N_{\rm sc}N_{\rm s})\) by reusing the total stream power.
For \(N=T_{\rm f}-T_0+1\ge1\) feasible intervals, cumulative
sums and checking \eqref{eq:rate_degradation} cost \(O(N)\).
The subsequent search uses \(O(\log(N+1))\) evaluations of \(G\)
when \eqref{eq:rate_degradation} holds and \(O(N)\) objective
evaluations otherwise. Thus, the scalar interval-selection complexity
and cumulative-sum storage are both \(O(N)\). Prediction,
transceiver computation, and interval selection are included
in the processing overhead.
\end{remark}

\section{Simulation Results}
\label{sec:evaluation}
In this section, the numerical study evaluates channel prediction, trajectory transfer, autoregressive prediction, and single-STA MIMO transmission. Link-level comparisons assess precoding, modulation and coding scheme (MCS) selection, and goodput under different CSI policies. Sounding-interval optimization performance is also analyzed in this section through simulation.

\subsection{Simulation Setup}

\begin{figure}[!t]
\centering
\noindent\makebox[0.93\linewidth][c]{%
\begingroup
\setlength{\unitlength}{\linewidth}
\begin{picture}(1,0.468)
\put(0,0){\includegraphics[width=\linewidth]{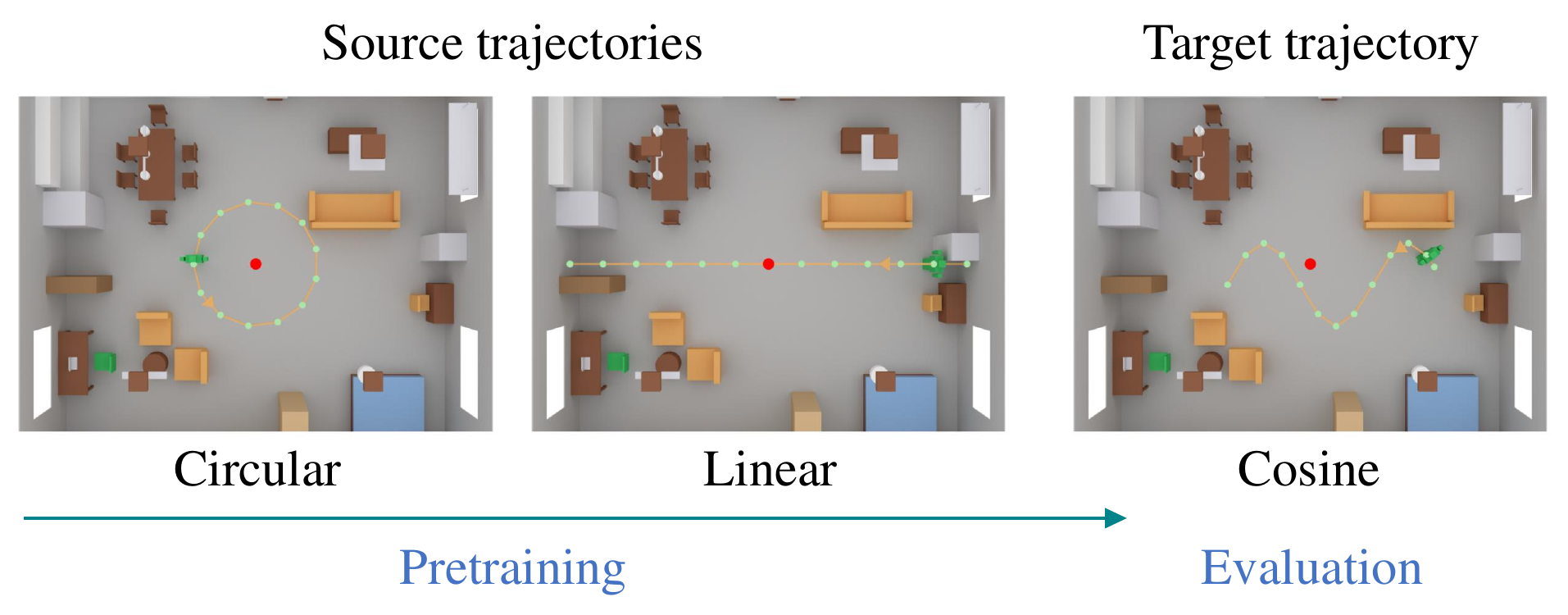}}
\end{picture}
\endgroup%
}\par
\caption{Linear and circular source trajectories and the cosine target trajectory in the simulation environment.}
\label{fig:transfer_scenario}
\end{figure}

The ray-tracing environment is a room containing concrete,
wood, plasterboard, metal, and etc., in which a fixed
AP serves a moving STA with various trajectories. At the stored snapshot \(m\), Sionna RT computes the MIMO channel
matrix \(\mathbf H_q[m]\) from the scene geometry, material
properties, and antenna configuration. Each CSI snapshot contains
\(242\times2\times8\) coefficients on the frequency grid defined in
Section~\ref{sec:system}, and the geometry is recomputed every
\(T_{\rm s}=0.1\) ms. For the per-step channel traces, the room includes a moving object; the spatial traces additionally include receiver coordinates. These settings are verified in Sionna RT 2.0.1 and the WLAN simulation is implemented with the aid of WLAN Toolbox Matlab on a Nvidia DGX Spark. Table~\ref{tab:sim_setup} summarizes the configuration. The source trajectories are linear and circular, and the transfer-learning target follows a cosine trajectory, as shown in
Fig.~\ref{fig:transfer_scenario}. The link simulations evaluate all
three trajectories.

\begin{table}[!t]
\centering\caption{Ray-Tracing and Evaluation Configuration}
\label{tab:sim_setup}\footnotesize
\begin{tabular}{@{}ll@{}}\toprule
Parameter & Value\\
Room dimensions & \(10\times10\times3.2\) $m^{3}$\\
AP position & \((5,5, 3.15, 1.35)\) m \\
Carrier / bandwidth parameter & 2.437 GHz / 20 MHz\\
Frequency samples / spacing & 242 / 78.125 kHz\\
Transmit / receive elements & 8 / 2\\
Maximum path depth / ray samples & 2 / \(5\times10^5\) per source\\
Channel-trace sampling interval & 0.1 ms\\
Circular radius / nominal speed & 1.5 m / 2 m/s\\
Displayed subcarrier index & 120\\
Channel samples / displayed duration & 1000 / 100 ms\\
Source / target window start index & 14500 / 23500\\
Delayed-CSI reference delay & 10 samples (1 ms)\\
Autoregressive prediction length & 30 steps\\
Linear / circular prediction start & 800 / 28000\\
\bottomrule\end{tabular}
\end{table}

\begin{figure}[!t]
\centering
\begin{minipage}[t]{0.48\linewidth}
\centering
\includegraphics[width=\linewidth]{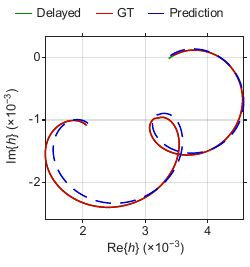}
\par{\footnotesize (a) Decoder-only 1 for RX1/TX2}
\end{minipage}\hfill
\begin{minipage}[t]{0.48\linewidth}
\centering
\includegraphics[width=\linewidth]{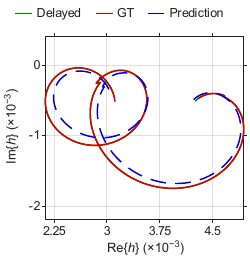}
\par{\footnotesize (b) Decoder-only 2 for RX1/TX7}
\end{minipage}\par
\begin{minipage}[t]{0.48\linewidth}
\centering
\includegraphics[width=\linewidth]{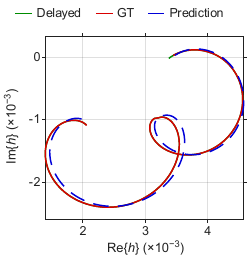}
\par{\footnotesize (c) RoPE 1 for RX1/TX2}
\end{minipage}\hfill
\begin{minipage}[t]{0.48\linewidth}
\centering
\includegraphics[width=\linewidth]{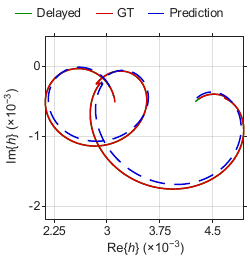}
\par{\footnotesize (d) RoPE 1 for RX1/TX7}
\end{minipage}\par
\begin{minipage}[t]{0.48\linewidth}
\centering
\includegraphics[width=\linewidth]{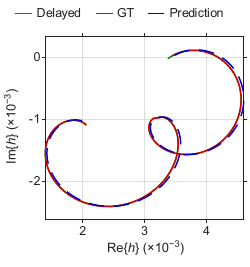}
\par{\footnotesize (e) RoPE + Huber 1 for RX1/TX2}
\end{minipage}\hfill
\begin{minipage}[t]{0.48\linewidth}
\centering
\includegraphics[width=\linewidth]{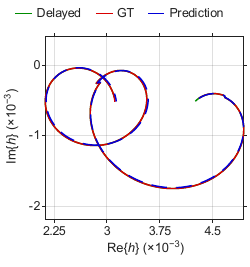}
\par{\footnotesize (f) RoPE + Huber 2 for RX1/TX7}
\end{minipage}
\caption{Ground-truth and predicted CSI:
(a)--(b) decoder-only; (c)--(d) RoPE; (e)--(f) RoPE+Huber.}
\label{fig:baseline_pairs}
\end{figure}

\begin{figure}[!t]
\centering
\noindent\makebox[\linewidth][c]{%
\includegraphics[width=0.48\linewidth]{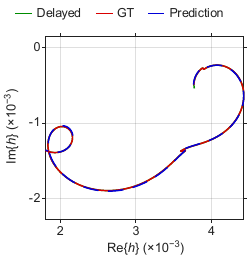}%
\hspace{0.04\linewidth}%
\includegraphics[width=0.48\linewidth]{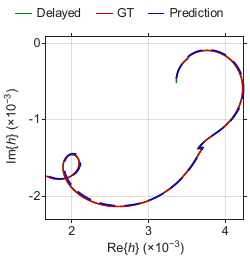}%
}\par
\caption{RoPE+Huber channel trajectories at RX0/TX1 and RX1/TX6.}
\label{fig:robust_pairs}
\end{figure}

\begin{figure}[!t]
\centering
\includegraphics[width=0.8\linewidth,height=0.48\linewidth]{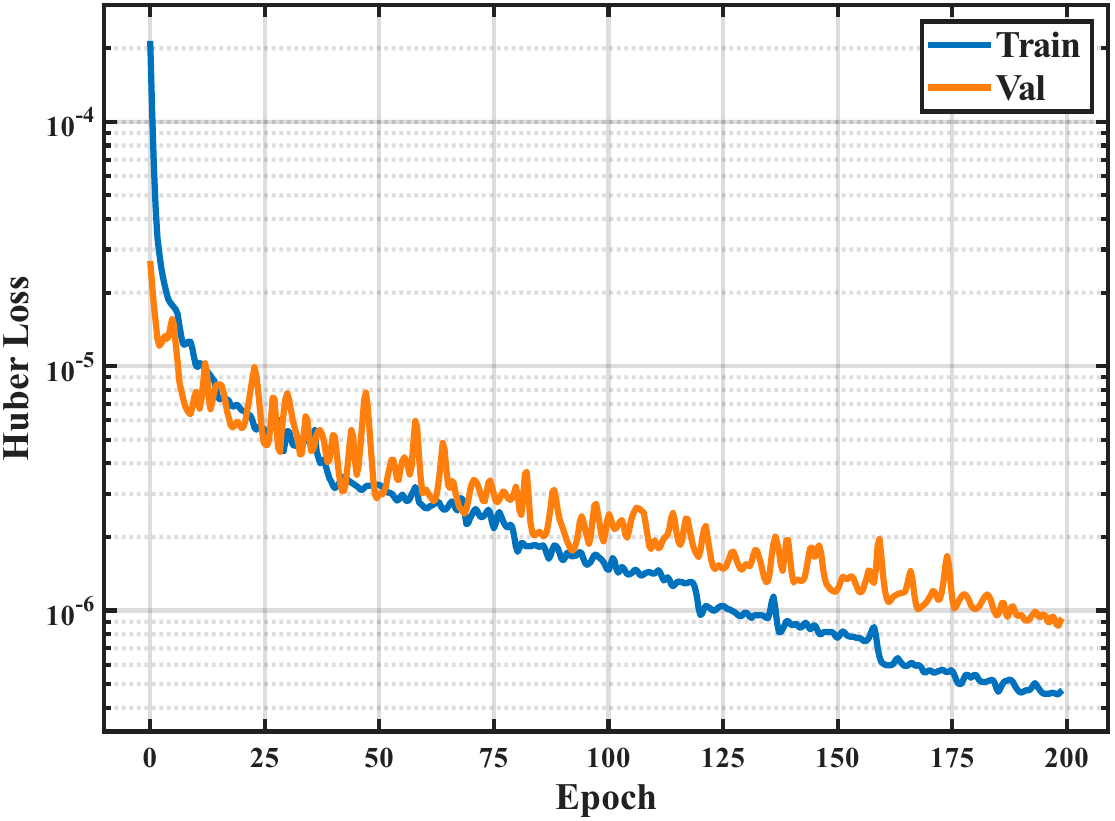}
\caption{Training and validation Huber loss of the model trained on the datasets for linear and circular trajectories.}
\label{fig:pretrain_loss}
\end{figure}

\begin{figure*}[!t]
\centering
\begin{minipage}[t]{0.28\textwidth}
\centering
\includegraphics[width=\linewidth]{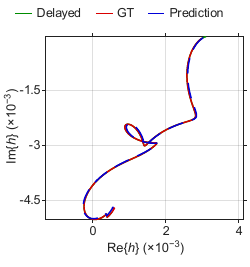}\par
\small (a) RX1/TX1
\end{minipage}\hfill%
\begin{minipage}[t]{0.28\textwidth}
\centering
\includegraphics[width=\linewidth]{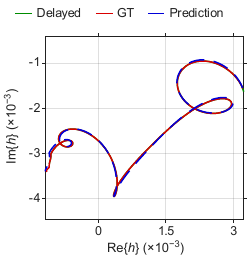}\par
\small (b) RX1/TX0
\end{minipage}\hfill%
\begin{minipage}[t]{0.28\textwidth}
\centering
\includegraphics[width=\linewidth]{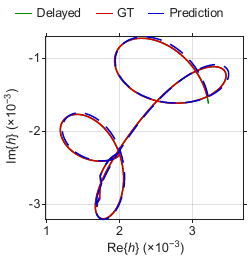}\par
\small (c) RX0/TX1
\end{minipage}
\caption{Zero-shot channel prediction on the cosine trajectory.}
\label{fig:wave_pairs}
\end{figure*}

\begin{figure*}[!t]
\centering
\begin{minipage}[t]{0.28\textwidth}
\centering
\includegraphics[width=\linewidth]{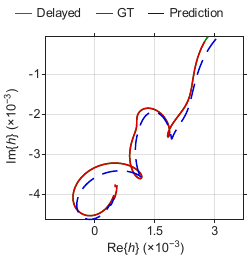}\par
\small (a) From scratch
\end{minipage}\hfill%
\begin{minipage}[t]{0.28\textwidth}
\centering
\includegraphics[width=\linewidth]{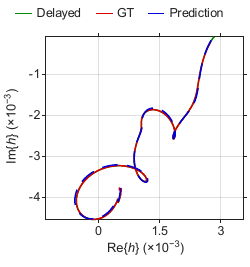}\par
\small (b) Zero-shot evaluation
\end{minipage}\hfill%
\begin{minipage}[t]{0.28\textwidth}
\centering
\includegraphics[width=\linewidth]{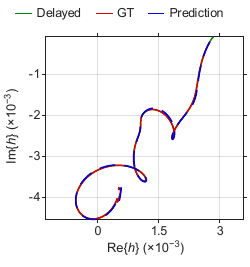}\par
\small (c) Transfer learning
\end{minipage}
\caption{Channel prediction for RX0/TX0 on the cosine target trajectory.}
\label{fig:transfer_pairs}
\end{figure*}

The prediction models share \(L=60\) history samples and
\(N_{\rm tar}=10\) training targets. Each token contains the
242 selected subcarriers of the 20-MHz channel, giving
\(D_0=2\times242\times2\times8=7744\). Each model has six decoder layers, embedding width \(d=128\), eight attention heads of dimension \(d_{\rm h}=16\), FFN width \(d_{\rm ff}=512\), and training dropout of 0.1. For each even-dimensional rotary component \(d'\), the frequencies are \(\omega_m=10000^{-2m/d'}\), \(m=0,\ldots,d'/2-1\).
The RoPE+Huber model utilizes the scaled Huber loss with \(\beta=10^{-3}\) for training:
\begin{equation*}
 \rho_\beta(z)=
 \begin{cases}
  \dfrac{z^2}{2\beta}, & |z|<\beta,\\[2pt]
  |z|-\dfrac{\beta}{2}, & |z|\ge\beta,
 \end{cases}
\end{equation*}
where \(z\in\R\) denotes the prediction error of a channel
feature after inverse normalization. The decoder-only and RoPE
models are trained with MSE. The optional spatial extension allocates \((d_{\rm t},d_{\rm x},d_{\rm y})=(8,4,4)\) coordinates to the temporal and two spatial rotation blocks. The reported channel-domain comparisons use temporal RoPE.
All models are trained with the Adam optimizer with a batch size of $8$ prediction steps that contain 60 frames as input and 10 frames as ground-truth output. Each element in the input frames comprises both CSI and the positional information of STA at a certain time instant. In each epoch, the forward propagation evaluates the predictions and the loss \(\mathcal L(\boldsymbol\theta)\) in \eqref{eq:training_problem} over the training set. Decoder-only, RoPE, and RoPE+Huber are each trained for 200 epochs with a learning rate \(3\times10^{-5}\), and training from scratch on the cosine trajectory runs 50 epochs with a learning rate \(3\times10^{-5}\). Transfer learning initializes all parameters from the pretrained RoPE+Huber model and first updates only the input projection \(\mathbf W_{\rm in},\mathbf b_{\rm in}\) and the prediction head \(\mathbf W_{\rm out},\mathbf b_{\rm out}\) in \eqref{eq:input_embedding}--\eqref{eq:predictor_io} for 15 epochs with a learning rate \(10^{-4}\), with the decoder layers frozen, i.e., their attention and FFN parameters in \eqref{eq:attention}--\eqref{eq:residual_updates} remain unchanged. Because the input projection and the prediction head connect the normalized CSI to the decoder layers, updating them adapts the model to the cosine trajectory, while the decoder layers keep their pretrained parameters unchanged. All parameters are then fine-tuned for 50 epochs with a learning rate \(3\times10^{-5}\), so that transfer learning and training from scratch update all parameters for the same number of epochs.

Let \(x_c^{\min}\) and \(x_c^{\max}\) denote the minimum and
maximum of the \(c\)th real or imaginary CSI component across
all channel snapshots in the dataset. Different CSI components can have different numerical ranges across subcarriers and antenna pairs. Those ranges affect how the components enter the learned projections and influence parameter updates. Feature-wise normalization gives the model a more consistent input scale, making optimization less sensitive to the raw CSI amplitudes. Hence, to place CSI components
on comparable scales, min--max normalization \(\mathcal N\) and its inverse are
\begin{align*}
 [\mathcal N(\mathbf X)]_{k,c}
 &=\frac{[\mathbf X]_{k,c}-\mu_c}{\lambda_c},\\
 [\mathcal N^{-1}(\mathbf X)]_{k,c}
 &=\lambda_c[\mathbf X]_{k,c}+\mu_c,
\end{align*}
where \(\mu_c=x_c^{\min}\) and
\(\lambda_c=x_c^{\max}-x_c^{\min}+10^{-8}\).
Each prediction step consists of 70 frames, 60 historical frames as input tokens to the Transformer with 10 output frames that comprise 10 predicted CSI frames separated by 0.1 ms; that is, 1 ms of predicted CSI with 0.1 ms granularity. Linear, circular, and cosine trajectories are sorted by start time. The first 80\% form the training set, and the remaining 20\% form the validation set. For the pretrained models, \(\mu_c\) and \(\lambda_c\) are computed from all linear and circular frames, with the training and validation Huber losses shown in Fig.~\ref{fig:pretrain_loss}. The training and validation Huber losses decline steadily over the full 200 epochs and the loss curves remain stable in the later training stage, indicating that the pretrained model achieves good convergence and exhibits no significant overfitting. Training from scratch computes \(\mu_c\) and \(\lambda_c\) from all cosine frames, and transfer learning reuses them stored in the pretrained models.

\subsection{CSI Prediction Performance}
Each comparison uses 1000 samples from all $16$ antenna
pairs. Decoder-only, RoPE, and RoPE+Huber are compared on a common source window. Zero-shot evaluation, training from scratch, and fine-tuning use a common cosine target window. The 100-ms traces concatenate one-step predictions obtained from updated ground-truth histories; autoregressive prediction is assessed separately over 30 steps. Let \(\mathcal T\) denote the 1000 evaluation indices and
\(\widehat{\mathbf H}[m]\in
\C^{N_{\rm r}\times N_{\rm t}}\) denote the predicted channel matrix. For non-zero reference energy, the
normalized MSE (NMSE) over all antenna pairs is
\begin{equation}
 \operatorname{NMSE}=
 \frac{\displaystyle\sum_{m\in\mathcal T}
 \|\mathbf H[m]-\widehat{\mathbf H}[m]\|_{\rm F}^2}
 {\displaystyle\sum_{m\in\mathcal T}
 \|\mathbf H[m]\|_{\rm F}^2}.
 \label{eq:test_nmse}
\end{equation}
For each antenna pair, NMSE is computed from its ground-truth
and predicted CSI. Figures~\ref{fig:baseline_pairs}--\ref{fig:transfer_pairs}
plot the ground-truth and predicted coefficients in the complex
plane in the time domain. The delayed-CSI baseline uses
\(\widehat{\mathbf H}^{(\mathrm{Delayed})}[m]
=\mathbf H[m-10]\)
over the 990 overlapping samples, corresponding to a delay
of \(10T_{\rm s}=1\) ms.
The link-level stale policy holds the most recently sounded
CSI until the next sounding update.

Figure~\ref{fig:baseline_pairs} shows decoder-only in (a)--(b),
RoPE in (c)--(d), and RoPE+Huber in (e)--(f).
Their NMSE values across all 16 antenna pairs are
\(5.1883\times10^{-4}\), \(4.1024\times10^{-4}\), and
\(4.2736\times10^{-5}\), respectively.
The 1-ms delayed-CSI baseline has an NMSE of
\(4.3233\times10^{-4}\).

RoPE+Huber achieves the lowest overall NMSE and outperforms the 1-ms delayed-CSI baseline on all 16 antenna pairs. RoPE and decoder-only outperform this baseline on 12 and 6 pairs, respectively.
In Figure~\ref{fig:baseline_pairs}(e)--(f), the RoPE+Huber predictions closely follow the ground-truth.
Figure~\ref{fig:robust_pairs} shows additional results for RX0/TX1 and RX1/TX6 in the left and right panels, respectively. This comparison evaluates the complete loss rather than the effect of Huber loss alone.

The cosine trajectory evaluates adaptation to a different motion pattern. Zero-shot prediction
directly uses the model trained on linear and circular trajectories. Transfer learning fine-tunes this model using cosine-trajectory data, whereas training from scratch uses random initialization. Figure~\ref{fig:wave_pairs}(a)--(c) show zero-shot predictions
for RX1/TX1, RX1/TX0, and RX0/TX1, respectively.
Across all 16 antenna pairs, the NMSE values are
\(1.2721\times10^{-4}\) for zero-shot prediction,
\(2.4279\times10^{-3}\) for training from scratch, and
\(7.8617\times10^{-5}\) for transfer learning, compared with
\(1.5237\times10^{-3}\) for the delayed-CSI baseline.
Zero-shot prediction reduces the overall NMSE by 91.65\%
relative to the delayed CSI and achieves lower NMSE on every
antenna pair. Transfer learning further reduces the overall
NMSE by 38.20\% relative to zero-shot prediction and by
96.76\% relative to training from scratch. Figure~\ref{fig:transfer_pairs}(a)--(c) compare training from
scratch, zero-shot prediction, and transfer learning for RX0/TX0.
For this antenna pair, the corresponding NMSE values are
\(2.7071\times10^{-3}\), \(8.1702\times10^{-5}\), and
\(4.8217\times10^{-5}\), respectively; the delayed-CSI
baseline has an NMSE of \(1.1820\times10^{-3}\).

\begin{figure}[!t]
\centering
\noindent\makebox[\linewidth][c]{%
\includegraphics[width=0.96\linewidth]{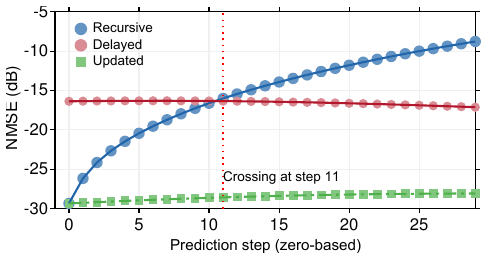}%
}\par
\noindent\makebox[\linewidth][c]{%
\includegraphics[width=0.96\linewidth]{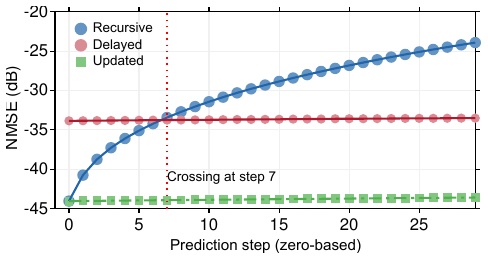}%
}\par
\caption{Autoregressive prediction residual NMSE for linear (top) and circular (bottom) motion. Updated represents the fresh ground-truth history. Delayed denotes the utilization of 1-ms-old CSI.}
\label{fig:iterative_drift}
\end{figure}

Figure~\ref{fig:iterative_drift} compares autoregressive prediction,
prediction using updated ground-truth histories, and 1-ms delayed
CSI for linear and circular motion.
Autoregressive prediction feeds each prediction back into the
history, whereas the updated method uses ground-truth CSI at every step.
Over 30 steps, the plotted autoregressive NMSE increases from
\(1.1625\times10^{-3}\) to \(1.3242\times10^{-1}\) for linear
motion and from \(3.9415\times10^{-5}\) to
\(4.0560\times10^{-3}\) for circular motion.
With updated ground-truth histories, the final NMSE values remain
at \(1.5562\times10^{-3}\) and \(4.3741\times10^{-5}\),
respectively.
The widening gap between the two prediction methods illustrates error accumulation when predictions are repeatedly reused: at these steps, autoregressive prediction becomes less accurate than simply reusing 1-ms-old CSI, showing that the accumulated feedback error outweighs the benefit of the learned model and motivating a finite prediction horizon and renewed CSI acquisition.

The growing gap between autoregressive and updated-history prediction
illustrates the cost of repeatedly feeding estimates back into the model.
An NMSE crossing can identify a loss of prediction advantage. It is noteworthy that the sounding decision may additionally depend on the per-stream SINR margin and
the airtime required for a new CSI acquisition.

\subsection{Link-Level Goodput With Predicted CSI}
\label{sec:link_goodput}

The link-level experiments compare ground-truth CSI, predicted CSI,
and stale CSI under the channel configuration in
Table~\ref{tab:sim_setup}. Ground-truth CSI uses the current channel,
predicted CSI uses the channel prediction, and stale CSI retains
the most recently sounded channel until the next update.
All three methods transmit through the same ground-truth channel
using two spatial streams and the same total transmit power.
The selected CSI determines SVD precoding and combining, as well
as MCS selection. The receiver estimates the CSI from noisy
packet training.

\begin{figure}[!t]
\centering
\noindent\makebox[\linewidth][c]{%
\includegraphics[width=\linewidth]{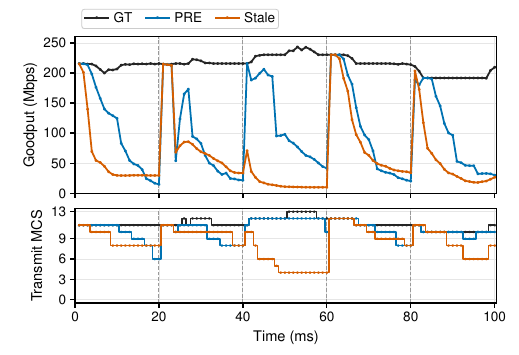}%
}\par
\caption{Goodput and transmit MCS for the linear trajectory.}
\label{fig:goodput_linear}
\end{figure}

\begin{figure}[!t]
\centering
\noindent\makebox[\linewidth][c]{%
\includegraphics[width=\linewidth]{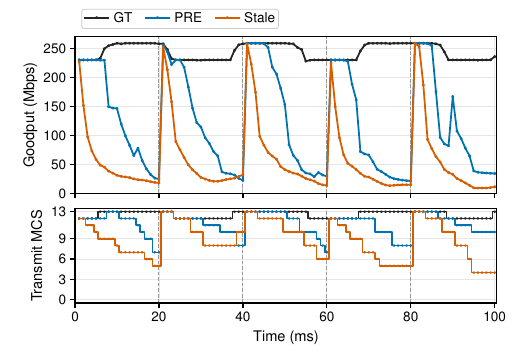}%
}\par
\caption{Goodput and transmit MCS for the circular trajectory.}
\label{fig:goodput_circular}
\end{figure}

\begin{figure}[!t]
\centering
\noindent\makebox[\linewidth][c]{%
\includegraphics[width=\linewidth]{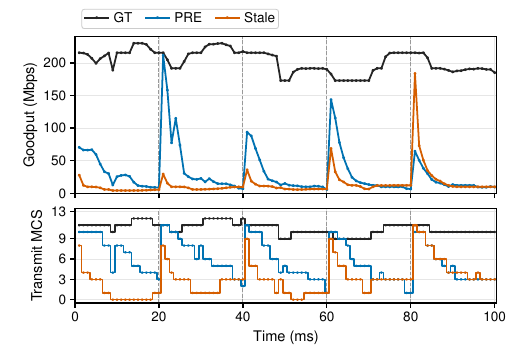}%
}\par
\caption{Goodput and transmit MCS for the cosine trajectory.}
\label{fig:goodput_cosine}
\end{figure}

For each method, goodput is calculated from the successfully
delivered payload bits \(D\) and total service airtime \(A\), measured in seconds: $G=D/A$ bits/s. Service airtime includes protocol overhead, failed transmission,
and retransmission. Mean goodput is the arithmetic average of 100 values evaluated at 1-ms intervals.
Figures~\ref{fig:goodput_linear}--\ref{fig:goodput_cosine}
show the resulting goodput and transmit MCS.

For the linear trajectory in Figure~\ref{fig:goodput_linear},
predicted CSI increases mean goodput from 61.25 Mb/s with stale
CSI to 102.47 Mb/s, with an improvement of 67.3\%.
Predicted CSI achieves higher goodput at 71 of the 100 evaluation
points. Ground-truth CSI yields a mean goodput of 214.94 Mb/s.

For the circular trajectory in Figure~\ref{fig:goodput_circular},
mean goodput increases from 58.12 to 123.64 Mb/s, with an improvement
of 112.7\%. This is the largest absolute gain among the three
trajectories, at 65.52 Mb/s. Predicted CSI outperforms stale CSI
at 97 of the 100 evaluation points, while ground-truth CSI
achieves a mean goodput of 244.89 Mb/s.

For the cosine trajectory in Figure~\ref{fig:goodput_cosine},
mean goodput increases from 13.93 to 30.52 Mb/s, yielding the
largest relative improvement of 119.1\%.
Despite this percentage gain, the absolute goodput remains below
that of the linear and circular trajectories.
Ground-truth CSI achieves a mean goodput of 202.67 Mb/s.

The benefit of prediction varies over time. Predicted CSI yields
lower goodput than stale CSI during 31--40 and 72--81 ms for
linear motion and 39--40 ms for circular motion.
During 72--81 ms in the linear trajectory, its selected MCS is
equal to or higher than that of stale CSI at eight of the ten
evaluation points, yet its goodput is lower.
A higher MCS therefore does not guarantee higher goodput:
successful decoding also depends on the channel mismatch
associated with the selected precoder.
These reversals motivate a finite prediction horizon and renewed
sounding, whose CSI acquisition overhead must be included when
selecting the sounding interval.

\begin{figure}[!t]
\centering
\includegraphics[width=0.85\columnwidth,height=0.65\columnwidth,keepaspectratio]{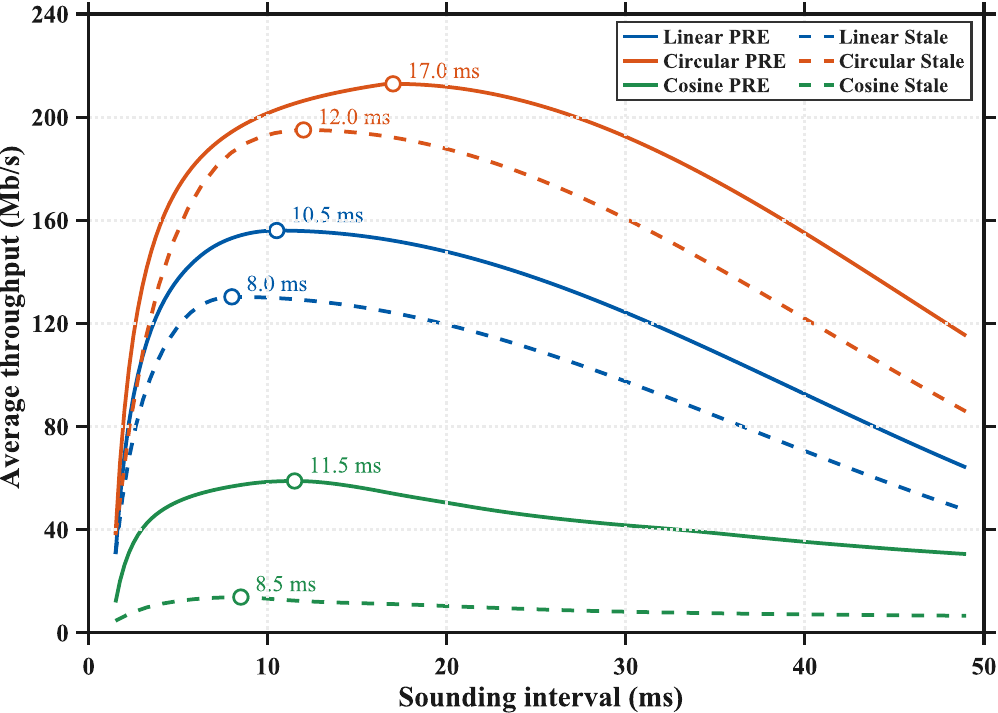}
\caption{Average throughput versus sounding interval.}
\label{fig:sounding_interval_illustration}
\end{figure}

Figure~\ref{fig:sounding_interval_illustration} schematically
illustrates the throughput--overhead tradeoff in \eqref{eq:utility}.
Shorter sounding intervals incur a larger overhead fraction,
whereas longer sounding intervals can increase channel mismatch and autoregressive
prediction error. The PRE curves illustrate the maxima at longer
intervals than their stale counterparts. 

\section{Conclusion}
\label{sec:conclusion}
This paper presented a prediction-aided channel sounding framework for time-varying MIMO Wi-Fi links. The framework connects autoregressive CSI prediction and prediction error to SINR-constrained interval selection, balancing throughput with sounding and processing overhead. The resulting predicted average-throughput maximization problem admits a globally optimal solution. In single-STA Wi-Fi 7 simulations, predicted-CSI beamforming improved mean goodput over stale CSI by 67.3\%, 112.7\%, and 119.1\% for linear, circular, and cosine trajectories, respectively. These results demonstrate the potential of CSI prediction to sustain MIMO link performance under receiver motion, while error accumulation in autoregressive prediction highlights the need for a finite prediction horizon.

\bibliographystyle{IEEEtran}
\bibliography{reference}
\end{document}


\appendix
\subsection{Throughput and Overhead Accounting}
\label{app:rate_coefficients}
Let \(N_{\rm sym}>0\) denote the number of data OFDM symbols per exchange, \(T_{\rm GI}\ge0\) denote the guard interval, and
\(T_{\rm MAC}\ge0\) denote the remaining exchange airtime,
excluding cycle-level sounding and processing.
With \(T_{\rm sym}=1/\Delta f+T_{\rm GI}\), the data-symbol
rate and first data slot are $c_{\rm r}=N_{\rm sym}/
 \{N_{\rm sym}T_{\rm sym}+T_{\rm MAC}\}$, and $T_0=1+\left\lceil
 (T_{\rm snd}+T_{\rm proc}) /\ T_{\rm s} \right\rceil$.
The timing parameters are fixed across candidate intervals,
and slots \(1,\ldots,T_0-1\) are fully reserved.
All controller inputs are available before processing, and
the common-horizon calculations are assumed complete before
slot \(T_0\).

\subsection{Proof of Theorem~1}
\label{app:sinr}
\begin{proof}
Fix a subcarrier and sample, and use the local SVD notation of
Section~III. Each combining row has an unit norm and precoding
column \(r\) has a norm \(\sqrt{p_r}\). For
\(\|\mathbf E\|_{\rm F}\le\varepsilon\),
\begin{align*}
 |e_{i,r}|&\le\sqrt{p_r}\|\mathbf E\|_2
 \le\varepsilon\sqrt{p_r},\\
 |g_i+e_{i,i}|&\ge[g_i-|e_{i,i}|]_+
 \ge[g_i-\varepsilon\sqrt{p_i}]_+,\\
 \sum_{\substack{r=1\\r\ne i}}^{N_{\rm s}}|e_{i,r}|^2
 &\le\varepsilon^2
 \sum_{\substack{r=1\\r\ne i}}^{N_{\rm s}}p_r.
\end{align*}
Here \(g_i=\sqrt{p_i}\widehat s_{q,i}[j]\ge0\).
The actual SINR denominator is positive because
\(\sigma_{\rm n}^2>0\). Applying the numerator lower bound
and denominator upper bound therefore gives
\begin{equation*}
 \gamma_i\ge
 \frac{[g_i-\varepsilon\sqrt{p_i}]_+^2}
 {\sigma_{\rm n}^2+\varepsilon^2
 \sum_{\substack{r=1\\r\ne i}}^{N_{\rm s}}p_r}
 =\gamma_i^{\rm lb}(\varepsilon).
\end{equation*}
For \(N_{\rm s}=1\), the empty interference sum is zero.
The argument applies to
every stream, subcarrier, and sample whenever their error bounds hold. No independence between the prediction error
and the singular vectors is required. The bound is sufficient
and can be conservative, since the separate desired-signal and
interference bounds need not be attained by the same error matrix.
\end{proof}

\subsection{Proof of Theorem~2}
\label{app:selection}
\begin{proof}
All candidates use the same predictions, powers, transceivers,
and error bounds. After Algorithm~1's first loop, the feasible
set is empty if \(T_{\rm f}<T_0\), and is
\(\{T_0,\ldots,T_{\rm f}\}\) otherwise: every longer candidate
contains a failed slot. In the empty case, Algorithm~1 returns
\(0\). Otherwise, its first loop computes
\(S[T]=\sum_{t=T_0}^{T}\widehat r[t]\) for every feasible endpoint.
Let \(T^*\) denote the smallest global maximizer of
\(\widehat R(T)=S[T]/T\) over this set. Direct subtraction gives
\begin{align*}
 \widehat R(T+1)-\widehat R(T)
 &=\frac{G(T)}{T(T+1)},\\
 \frac{G(T+1)-G(T)}{T+1}
 &=\widehat r[T+2]-\widehat r[T+1].
\end{align*}
The identities hold for \(T_0\le T<T_{\rm f}\) and
\(T_0\le T<T_{\rm f}-1\), respectively.
Under the rate-degradation condition, \(G(T)\) is non-increasing.
Hence throughput strictly increases before the first
\(G(T)\le0\) and does not increase afterwards. That endpoint
is \(T^*\); if every \(G(T)>0\), then \(T^*=T_{\rm f}\).
For a singleton feasible set, \(T^*=T_0=T_{\rm f}\).

Binary search maintains \(a\le T^*\le b\), initialized by
\(a=T_0\) and \(b=T_{\rm f}\). Since \(m<b\le T_{\rm f}\),
\(G(m)\) is defined. If \(G(m)\le0\), then \(T^*\le m\);
otherwise, \(T^*>m\). Each update preserves the invariant
and strictly reduces \(b-a\), so termination gives
\(a=b=T^*\). If the rate-degradation condition fails,
Algorithm~1 compares every feasible average; strict updates
retain the smallest global maximum. Finally, Theorem~1 gives
\(\gamma_{q,i,t}\ge\gamma_{q,i,t}^{\rm lb}\ge\gamma_{\min}\)
throughout the selected non-zero interval whenever the channel
errors satisfy their bounds.
\end{proof}